\pdfoutput=1
\documentclass[fleqn,usenatbib]{mnras}

\usepackage[T1]{fontenc}
\usepackage{ae,aecompl}

\usepackage{graphicx}	
\usepackage{amsmath}	
\usepackage{amssymb}	
\usepackage[utf8]{inputenc} 
\usepackage{hyperref}

\def\m2s2{\hbox{\,m$^{2}$\,s$^{-2}$}} 

\def\Mjup{\hbox{$\mathrm{M}_{\rm Jup}$}}
\def\Mearth{\hbox{$\mathrm{M}_{\oplus}$}}
\def\Rearth{\hbox{$\mathrm{R}_{\oplus}$}}

\def \1s{$1\,\sigma$}

\def \t0{T$_0$}

\title[An upper boundary in the mass-metallicity plane of exo-Neptunes]{An upper boundary in the mass-metallicity plane of exo-Neptunes}

\author[B. Courcol et al.]{
Bastien Courcol,$^{1}$\thanks{E-mail: bastien.courcol@lam.fr}
François Bouchy,$^{1}$
and Magali Deleuil$^{1}$
\\
$^{1}$Aix Marseille University, CNRS, Laboratoire d'Astrophysique de Marseille UMR 7326, 13388 Marseille cedex 13, France\\
}

\date{Accepted XXX. Received YYY; in original form ZZZ}

\pubyear{2016}

\begin{document}
\label{firstpage}
\pagerange{\pageref{firstpage}--\pageref{lastpage}}
\maketitle

\begin{abstract}
With the progress of detection techniques, the number of low-mass and small-size exoplanets is increasing rapidly. However their characteristics and formation mechanisms are not yet fully understood. The metallicity of the host star is a critical parameter in such processes and can impact the occurence rate or physical properties of these planets. While a frequency-metallicity correlation has been found for giant planets, this is still an ongoing debate for their smaller counterparts. Using the published parameters of a sample of 157 exoplanets lighter than 40 {\Mearth}, we explore the mass-metallicity space of Neptunes and Super-Earths. We show the existence of a maximal mass that increases with metallicity, that also depends on the period of these planets. This seems to favor in situ formation or alternatively a metallicity-driven migration mechanism. It also suggests that the frequency of Neptunes (between 10 and 40 {\Mearth}) is, like giant planets, correlated with the host star metallicity, whereas no correlation is found for Super-Earths (<10 {\Mearth}).
\end{abstract}

\begin{keywords}
Planetary Systems, planets and satellites: terrestrial planets -- Planetary Systems, methods: statistical -- Astronomical instrumentation, methods, and techniques
\end{keywords}



\section{Introduction}

Thanks to continuous improvements of detection techniques, the list of known exoplanets has exponentially increased these twenty past years. More specifically a new population of low-mass or small-size exoplanets has emerged both from the high-precision radial velocity surveys \citep{Mayor2009, Howard2009} and high-precision photometric surveys \citep{Baglin2003, Borucki2011}. Arbitrarily this population usually distinguishes Neptune like objects, with a mass from 10 to 40 {\Mearth} and a radius from 2 to 6 {\Rearth}, and super-Earths, with a mass from 2 to 10 {\Mearth} and a radius smaller than 2 {\Rearth}. To investigate the properties of the different planet populations in the low mass regime, and explore possible correlation between their physical properties, it is however required to get a large sample of planets with parameters accurately determined. \\
\indent Among the properties of exoplanet host stars, the metallicity was early identified as a key element for giant gaseous planets.  It has been well established that the occurrence rate of giant planets increases with metallicity, e.g. \cite{Gonzalez1997, Laws2003, Santos2005, Sousa2008}. Nevertheless, for low-mass exoplanets such a correlation was not observed (.e.g.  \cite{Sousa2008, Ghezzi2010, Mayor2011, Sousa2011}). \cite{Jenkins2013} claim the existence of a minimal mass for super-Earth objects that increases with host star metallicity. More recently \cite{Wang2015}, based on Kepler results, pointed out a universal correlation between the occurrence rate and the host star metallicity, which is weaker for terrestrial planets.  \\
\indent We took advantage of the increasing number of low-mass exoplanets published so far to explore and better quantify the possible correlation between their mass and the metallicity of their host star. Section 2 presents the sample of low mass planets we used. In section 3, we study the mass-metallicity diagram and show the existence of a upper boundary in the mass-metallicity plane as well as investigate its correlation with the period. Section 4 details the impact of this mass-metallicity trend on the frequency of low-mass planets. We discuss our findings in Section 5 and finally present our conclusions in the Section 6.  

\section{The sample} \label{sample}

The sample contains all the known low-mass exoplanets ($Msin(i) < 40\Mearth$) with a precision on the measured mass better than 20\% and a precision on the metallicity index [Fe/H] better than 0.2 dex. It was built on the basis of the main websites exoplanets catalogs : the NASA Exoplanet Archive \footnote{http://exoplanetarchive.ipac.caltech.edu/index.html}, exoplanets.org and exoplanet.eu. The metallicities, planetary masses and other parameters were carefully cross-checked between catalogs. We also included 44 planets from \cite{Mayor2011} as they are already present in the exoplanet.eu database, updated by the recent resubmitted version (private communication). This new set of low-mass planets increases the size of the sample by $27\%$ but do not change nor impact our results. \\
\indent The final list with references is displayed in Annex tables. It contains 157 planets with masses and metallicities ranging from 1.13 {\Mearth} to 38.1 {\Mearth} and from -0.89 to 0.39 dex respectively. The stellar type of the host stars range from M to F. We note that 88\% of the planets have periods less than 100 days. Twenty five planets were detected in transit, including some planets with masses determined by transit timing variations (TTVs), emphasizing the small overlap between radial velocity and transit surveys.\\
\indent Due to the multiplicity of the sources, there is no uniform metallicity determination method. Moreover, in some cases, the results of the different methods wildly disagree, with differences that can reach $\sim$0.3 dex \citep{Johnson2009, Neves2012}. To mitigate that effect, that could induce biases, we used the metallicity values from SWEET-Cat \citep{SWEET} whenever possible. This catalog aims at determining atmospheric parameters of exoplanet host stars in the most uniform way possible using the same methodology as well as compile values in the literature in a way that optimizes the uniformity, making them more suitable for statistical studies of stars with planets. In the present case 134 of our 157 planets, in 80 of the 97 systems, are present in the catalog. We note that the remaining planets are quite uniformly distributed in the parameters space and should not introduce any significant bias.

\section{The mass-metallicity diagram}

\subsection{On the existence of an exclusion zone}

\begin{figure}
   \centering
   \includegraphics[width=\hsize]{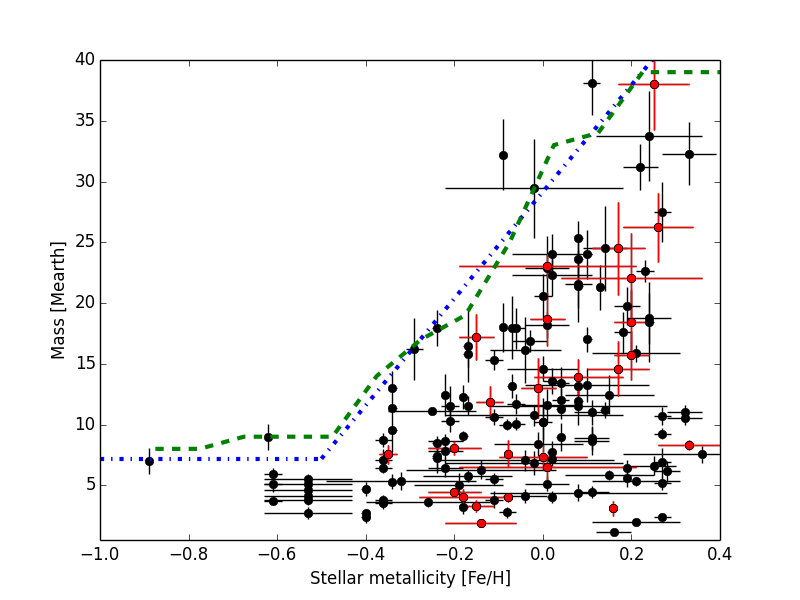}
      \caption{Planetary mass / host star metallicity diagram for all known planets lighter than 40 {\Mearth}, with $\Delta M/M < 0.2$ and $\Delta \mathrm{[Fe/H]} <0.2$ dex. Red dots are transiting planets. Some errors bars do not appear either because errors were not provided or because it is below the size of the dot. The dashed green line is the computed boundary, and the dashed blue line its approximation described in equation \ref{equation} (see text).}
         \label{Femass}
\end{figure}

\begin{figure}
   \centering
   \includegraphics[width=7cm]{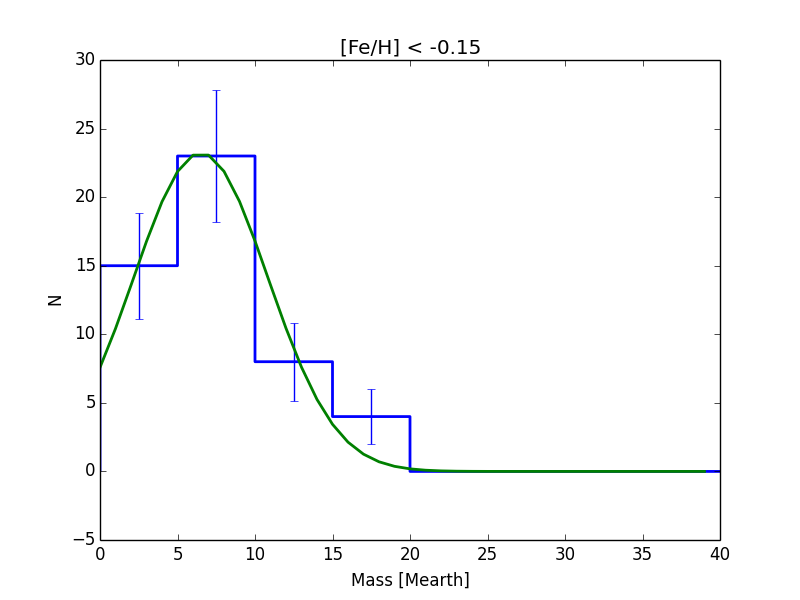}   
   \includegraphics[width=7cm]{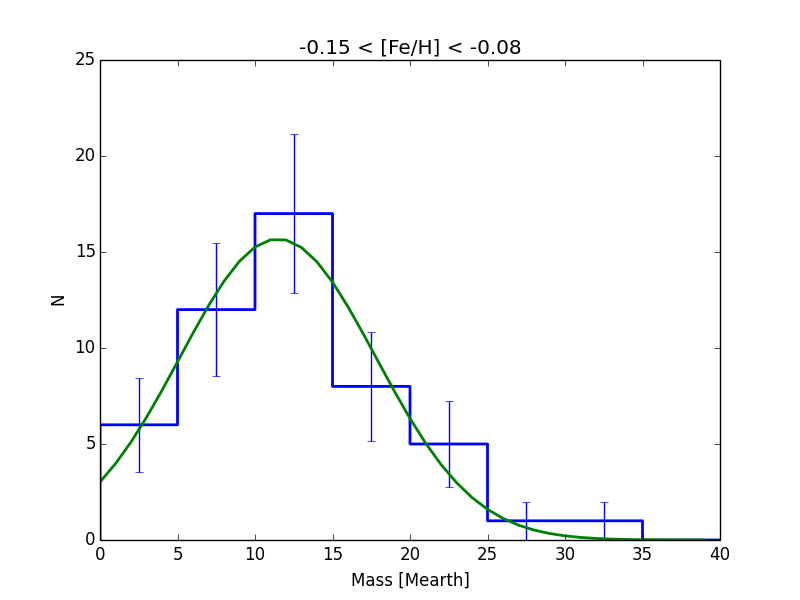}  
   \includegraphics[width=7cm]{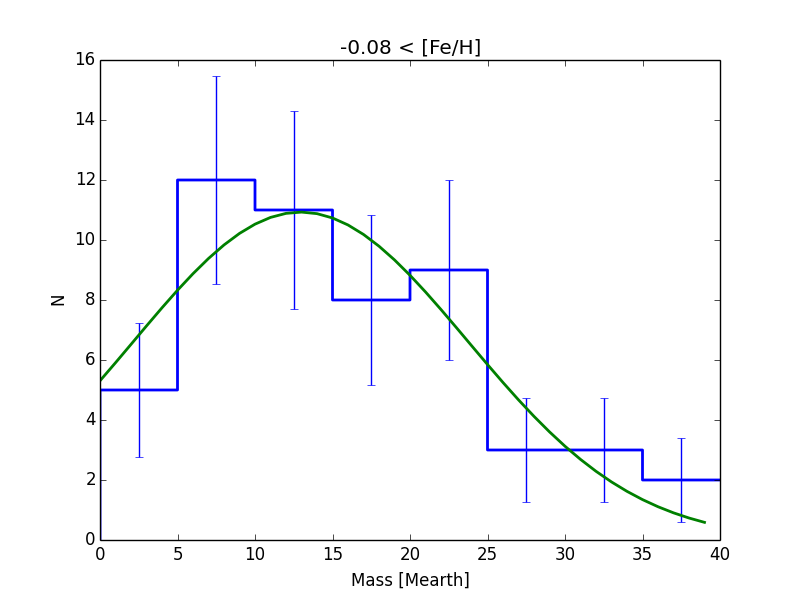}  
      \caption{Mass distributions for metallicity bins of $\sim50$ planets (blue histograms) and the corresponding gaussian fits (green curves). The metallicity range of the bins and the parameters of the gaussian fits (mean $\overline{M}$ and dispersion $\sigma_M$) are, from top to bottom : -0.89 to -0.15 dex, $\overline{M}=6.4$ {\Mearth}, $\sigma_M=4.2$ {\Mearth} ; -0.15 to 0.04 dex, $\overline{M}=11.6$ {\Mearth}, $\sigma_M=6.2$ {\Mearth} ; -0.04 to 0.39 dex, $\overline{M}=13.7$ {\Mearth}, $\sigma_M=11.2$ {\Mearth}. The errors bars correspond to a Poisson noise.}
         \label{distri_bins}
\end{figure}

\indent The planetary minimal mass/host star metallicity diagram is presented in Figure \ref{Femass}. While a connection between mass and metallicity is obvious, to describe it as a correlation would be misleading as the mass does not necessarily increase with the host star metallicity. A more adequate description would be that there is a maximum mass that increases with metallicity i.e. an upper boundary, or that there is an exclusion zone in the upper left part of the Figure \ref{Femass}, that is high masses and low metallicity.\\
\indent This type of dependency between two parameters is unusual and might be interpreted at first glance by a bias in the sample, as the dispersion of the mass increases with metallicity. First, biases related to the stellar type should be reviewed. Low mass planets are more easily detected around M dwarves because of their lower masses. Additionnaly, the precise determination of the metallicity of these stars is more difficult. However, when removing all the M dwarves from the data, the shape of the exclusion zone remains the same. More generally, as explaned in section \ref{sample}, the use of SWEET-Cat should prevent any bias caused by different metallicity determination methods. \\
\indent Furthermore, such a bias cannot be observational in nature, as it is the low-mass planets that are the hardest to detect that are found at low metallicity. Small planets orbiting low-metallicity host stars are the most difficult to detect, because of the much lower number of spectral features that can be exploited to obtain a precise radial velocity measurement. Furthermore, if more massive planets would have existed in these systems, they should have been detected. \\ 
\indent This exclusion region cannot be explained either by a bias in the angle $i$ between the plane of the system and the line of sight, which is unknown for most planets in this sample. The distribution of $i$ is a purely geometrical effect that is not linked to the stellar metallicity. We also note that the number of transiting planets for which the true mass is known is small in this mass domain. However, from a statistical point of view, the use of the minimal mass instead of the true mass should not significantly change the shape of this distribution, but only raise it by $\sim$15\%. For this reason, and for the sake of simplicity, the general term 'mass' will be used hereafter instead of 'minimal mass' (or 'true mass'), except where a distinction is needed. \\
\indent The only possible type of bias would then come from differences in the completeness in period of surveys focusing on either end of the metallicity range. If the periods probed around low metallicity stars are significantly shorter than high metallicity stars, it could explain the trend in Figure \ref{Femass} if Neptune-like planets are preferentially located at longer periods. However that cannot be the case, because the periods of the planets are not correlated with the metallicity. We discuss further this point in section \ref{per}. Additionnally, we could set up a strict period criterium to ensure a homogeneous completeness of the sample. If we discard all the planets at periods greater than 10 days (95), the general shape of the mass/metallicity diagram in Figure \ref{Femass} does not change. \\ 
\indent We finally explored the underlying population distribution in different [Fe/H] bins. To that purpose we divided our sample in 3 sub-samples of increasing metallicity with approximately the same number of planets. Assuming a poissonnian noise, we checked that each sub-sample can be phenomenogically described by a single gaussian function within the error bars (cf Figure \ref{distri_bins}). This is an additional evidence that biases are not correlated with [Fe/H] and therefore could not explain the exclusion zone. We also note that, as expected, the mean and the dispersion of the gaussians increase with the metallicity.\\

\subsection{Determination of the upper boundary}

\indent It is possible to define a mass-metallicity boundary separating the planets from the exclusion zone. To determine its shape we computed the cumulative distribution of planetary masses over a succession of metallicity bins. To account for the error bars and the possibility of outliers, each mass is weighted by the inverse of its precision. The "maximum mass" of the bin is set as the 97\% limit of this cumulative weighted distribution. The extremum bins (at -0.9 and 0.4 dex) are set to the closest bin value to overcome boundary effects. The green dashed line on Figure \ref{Femass} is the limit derived with this method with bins centered every 0.1 dex and 0.25 dex wide (therefore overlapping, to smooth the limit). Modifying the bin size and spacing can slightly change the shape of the boundary, without any significant impact. \\
\indent The main characteristic of the boundary is a monotonous increase of the maximal mass with [Fe/H]. For metallicities above -0.5, the trend seems linear. We therefore performed a linear regression on the boundary to get a simple relation approximating $M_{\mathrm{max}}$ (in {\Mearth}) as a function of [Fe/H] in equation \ref{equation}. \\

\begin{equation}\label{equation}
\mathrm{[Fe/H]}>-0.5  : M_{\mathrm{max}} = 43.3 \times \mathrm{[Fe/H]} + 29.2 {\Mearth}
\end{equation}

For metallicities below -0.5, the boundary is rather flat. However the reality of this plateau is questionable as it relies only on one peculiar planet, Kapteyn’s c \citep{Anglada2014}. Its host is an old sub-M dwarf of the halo, the only one of the sample, which has a singular metallicity of -0.89 dex with no associated error. Moreover the orbital period of Kapteyn's c is among the longest of the sample: 121.5 days. It is possible that the planetary properties of such systems are different and that the flat trend is not representative of the global population.\\

	\subsection{Correlation with the period}\label{per}

\begin{figure}
   \centering
   \includegraphics[width=\hsize]{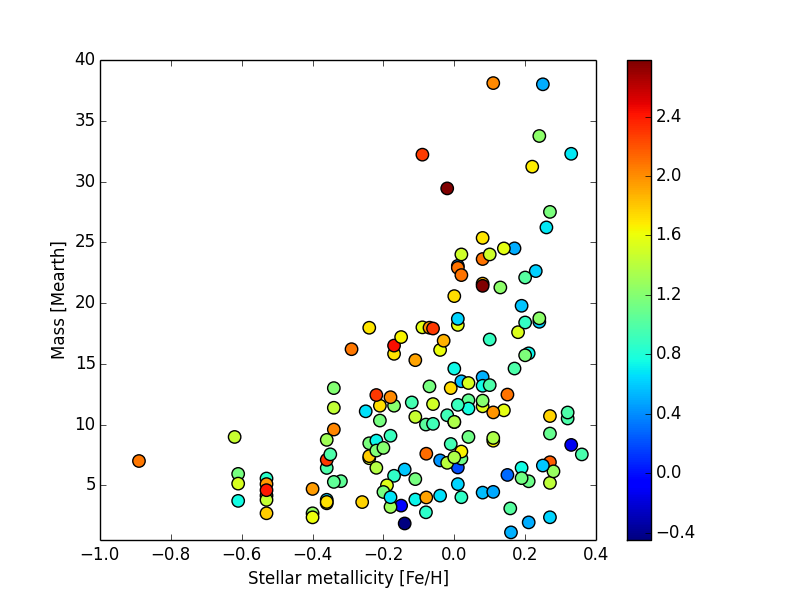}   
   \includegraphics[width=\hsize]{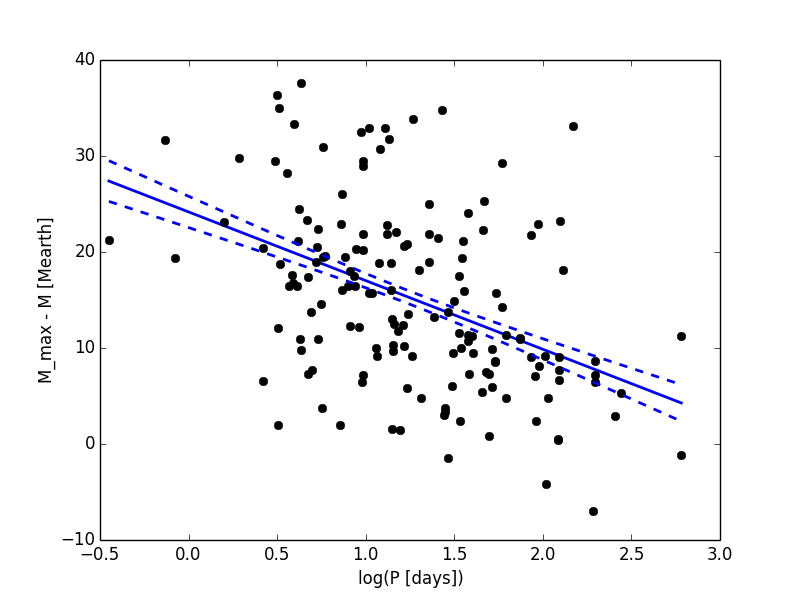}
      \caption{\textit{Upper panel} : distribution of the sample in the mass/metallicity plane with $\log{P}$ as the color scale. \textit{Bottom panel} : $\log{P}$ versus $M_{\mathrm{max}}-M_{\mathrm{planet}}$, $M_{\mathrm{max}}$ computed from equation \ref{equation}. The solid blue line is the linear regression of the data and the dashed blue lines are the 1 $\sigma$ confidence interval.}
         \label{Femass_per}
\end{figure}

The unusual nature of this connection between the planetary mass and the stellar metallicity could be explained by the existence of correlations with other parameters. We therefore investigate a possible correlation with the orbital period. The Figure \ref{Femass_per}, top panel, represents the mass-metallicity plane, with the logarithm of the period as the color scale. We notice that the planets closer to the limit tend to have longer period compared to those farther away. Moreover, the few planets that are slightly above the defined boundary all have periods greater than 100 days. This is also the case of Kapteyn's c, which is responsible of the questionable plateau for extremely low metallicities. This can be best seen in the bottom panel, which represents $\log{P}$ versus $M_{\mathrm{max}}-M_{\mathrm{planet}}$, i.e. the vertical distance to the boundary as defined by the equation \ref{equation} (although we did not allow $M_{\mathrm{max}}$ to increase further than 40 {\Mearth} as it is the limit of our sample). When performing a linear regression in the data (the blue dotted line in the bottom panel), we obtain the following relation: 

\begin{equation}\label{equation_per}
M_{\mathrm{max}}-M_{\mathrm{planet}} = -7.15 \pm 1.17 \times \log{P} + 24.17 \pm 1.6 {\Mearth}
\end{equation}

For this regression we used an identical weight for all the planets. Indeed, in this sample, the uncertainties on the mass are not homogeneously computed. Moreover, other sources of uncertainties should be taken into account in the error on the distance to the boundary (e.g. the uncertainty introduced by the sin(i), the error on the position of the limit) that are beyond the scope of this paper. There is consequently no solid argument to give more weight to some planets. \\
\indent The parameters of the linear regression are significantly constrained (3.08 $\sigma$ for the slope), although the dispersion of the residuals is quite high. Similarly, the Pearson correlation coefficient of this data set is of -0.44, but the probability of the no-correlation hypothesis (P-value) is of 8e-9. This means that while the correlation between $\log(P)$ and $M_{\mathrm{max}}-M_{\mathrm{planet}}$ is weak, it is very significative. \\
\indent This result indicates the upper limit decreases for short period planets. More interestingly, this also suggests that Neptune-like planets could still exist around metal-poor stars, but at longer periods. 

	\subsection{On the existence of a lower boundary}

\begin{figure}
   \centering
   \includegraphics[width=\hsize]{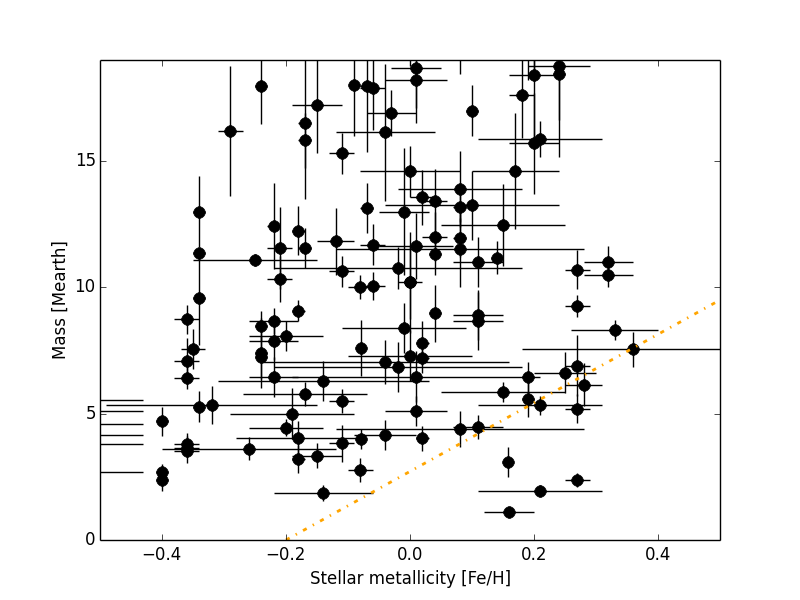}   
      \caption{Mass-Metallicity distribution of known small planets. The orange dashed line is the lower boundary proposed by Jenkins et al. 2013.}
         \label{jenkins}
\end{figure}

\indent \cite{Jenkins2013} also discussed of the possible correlations between mass and metallicity for low mass planets. Their study, based on the exoplanets.org database as of 2012, focused on a smaller range of masses (0 to 19 {\Mearth}) and metallicity (-0.5 dex to 0.5 dex). Instead of a correlation, they proposed the existence of a lower boundary, increasing linearly from 0 {\Mearth} at -0.2 dex to 9.5 {\Mearth} at 0.5 dex.\\
\indent With a sample tripled in size, 121 planets in our study instead of 36 in Jenkins et al 2013 in the same mass-metallicity range, it is possible to test this boundary with a better reliability. Our sample is represented in Figure \ref{jenkins} in their mass-metallicity range. Eight planets are found below this boundary (the orange dashed line), 3 of them (alpha Cen B b, HD 134606 b, GJ 876 d) at more than 1 $\sigma$ if we consider conservative errors in [Fe/H] of at least 0.1 dex. This boundary therefore does not hold up when faced to new detections. 

\section{Impact on the frequency of small planets} \label{freq}

\begin{figure}
   \centering
   \includegraphics[width=\hsize]{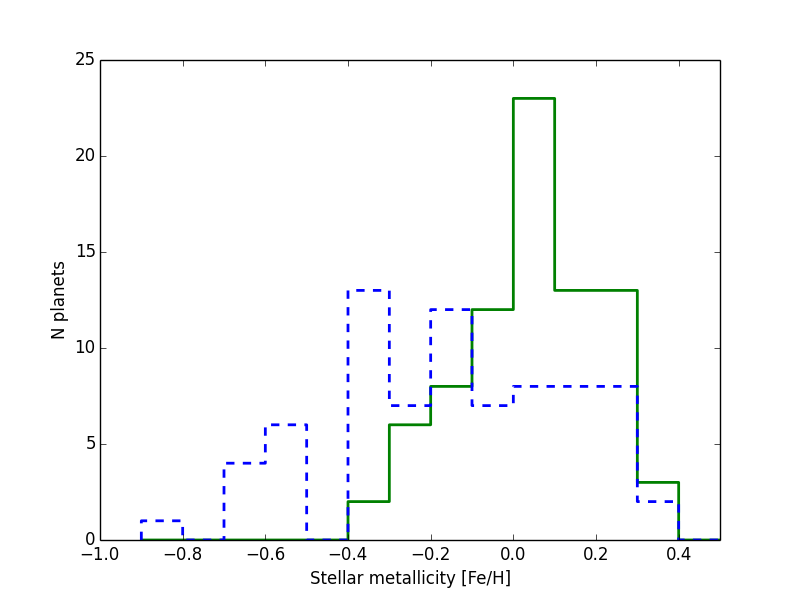}   
      \caption{Number of the Neptune-mass planets (between 10 and 40 {\Mearth}) (green solid line) and Super-Earth planets (<10 {\Mearth}) (blue dotted line). Interestingly, the number of planets in each category is similar (80 Neptunes and 75 Super-Earth).}
         \label{distrib}
\end{figure}

\indent The limit between Neptune and Super-Earths is difficult to place. Currently, no clear mass criterium exists to discriminate these two populations, either observationnally or physically motivated. This can be explained by three reasons. First, degeneracies in planet interiors models make the categorization uncertain for a range of scenarios. Second, the existence of transitional planets (Mini-Neptunes, Mega-Earths or low density Super-Earths) can further scramble the limit to an extent that is currently unknown. And finally, there is a lack of observational constrains with only 25 small planets with a relatively well measured density. Consequently we can only choose an arbitrary mass to define "Neptune-mass" and "Super-Earth" planets, in this case the widely used 10 {\Mearth}. However, we can already perceive different statistical behaviors with the current sample. \\
\indent Figure \ref{distrib}, shows the number of "Neptune-mass" planets (between 10 and 40 {\Mearth}) as a function of the metallicity (green solid curve). The shape of this distribution cannot be used as an occurence rate, as the planets come from many sources with different selection biases, methods and detection performances. Nonetheless, it gives an important clue. One can see that the distribution drops to 0 at -0.4 dex for the whole sample (green curve), which is coherent with the shape of the exclusion zone. This implies that at the first order, the frequency of Neptunes is correlated to metallicity, with no Neptunes around very sub-metallic stars. Beyond that observation, it would not be surprising that the frequency of Neptunes do follow the observed distribution at low metallicities and steadily decrease to reach 0. Super-Earths (smaller than 10 {\Mearth}, blue dotted curve in Figure \ref{distrib}) on the other hand, are present at all metallicities. We note that this is still true for all limit masses used to define Neptunes and Super-Earths between 8 and 15 {\Mearth}. \\
\indent This result differ from previous studies \citep{Ghezzi2010, Mayor2011, Sousa2011}. In the study of 582 FGK stars of a HARPS volume limited subsample of \cite{Sousa2011}, the metallicity of Neptunian hosts is rather flat compared to that of the stars hosting jovians, although they could not achieve statistically meaningful results due to small numbers. \cite{Mayor2011} present a flat, metal poor (<0.2 dex besides one exception) distribution for planets less massive that 30 {\Mearth}. Finally \cite{Ghezzi2010} also found a flat relative frequency for Neptune-mass hosts in regard to the metallicity, even when adding planets from the literature to their results. Complementary to these observational results, \cite{Mordasini2012} reported no correlation between the protoplanetary disk metallicity and the frequency of Neptunes using formation models.\\
\indent Two factor may explain this discrepancy. First, the number of planets used in our study is much larger and therefore more statistically reliable. Second, the previous studies on this matter seldom make the distinction between Neptunes and Super-Earths, and sometimes define "Neptunes" as any planet with a mass lower than a given value (25 {\Mearth} in \citealt{Ghezzi2010}, 30 {\Mearth} in \citealt{Mayor2011}, 0.1 {\Mjup} in \citealt{Sousa2011}). This is important considering that the distribution of Super-Earths (smaller than 10 {\Mearth}) is different to that of the Neptunes (between 10 and 40 {\Mearth}), and dilute the significance of any trend visible for Neptunes only. \\
\indent Our results are also in good agreement with the more recent paper from \cite{Wang2015}. These authors analyzed the Kepler results and pointed a universal correlation between the occurence rate and the host star metallicity. They show that this dependancy is weaker for terrestrial planets than for gas-dwarf planets. This is consistent with the fact that we can extrapolate such a correlation for our Neptune sample but not for our Super-Earth sample, while both are roughly equal in size. However we note that they use photometrically derived KIC metallicities, that have a poor precision.

\section{Discussion}

While the observed correlation does not imply causation, it is interesting to consider the possibility that the metallicity would drive the maximal mass of these small planets. A mass-metallicity upper boundary fits well within the core-accretion theory \citep{Pollack1996}. Indeed, a metal poor star could see its planets forming smaller cores, that will thus accrete gas less efficiently. This translates as a maximum mass rather than a strict correlation probably because of other phenomena, such as the competition between multiple planets that decrease the quantity of matter available for a single one.\\
\indent The correlation between this boundary and the period is also peculiar. A possible explanation would be that the highly irradiated parts of a protoplanetary disk are more depleted of volatile elements due to intense radiation pressure. The planets forming in these regions have a much lower amour of gas and rely more exclusively on the presence of heavier elements. Such a scenario implies that these planets are formed in situ, or at least that a significant fraction of their total mass is acquired during or after the migration. Another possible explanation is that planets do form at all metallicities, but that in metal-poor disks Neptune-like planets form farther out or do not migrate as quickly as in metal-rich disks. These planets would consequently lie outside the period ranges that are probed by current surveys. This is the mechanism proposed by \cite{Adibekyan2013}, who showed that the periods of planets more massive than 10 {\Mearth} orbiting metal-poor stars are preferentially longer than those orbiting metal-rich stars. However, the role of other parameters like the mass of the protoplanetary disk, the number of planets and architecture of the system or the migration type should be taken into account. \\
\indent Similar studies have been performed on the Kepler Objects of Interest sample when considering the radius instead of the mass. \cite{Buchhave2014} show that the mean host star metallicity is statistically higher for larger planets, with spectroscopically derived metallicities. They categorize them into three groups of increasing metallicities : the terrestrial planets $R_P < 1.7 R_\oplus$, gas dwarf planets $1.7 R_\oplus < R_p < 3.9 R_\oplus$ and gas or ice giants $3.9 R_\oplus < R_p$. These results are consistent with the trend we observe for the masses. \\
\indent Based on the same metallicity values, \cite{Dawson2015} go one step further and show that for period greater than 15 days, for which there is no significant photoevaporation, there is a lack of rocky planets ($R_P < 1.5 R_\oplus$) around metal rich star, while all types of planets are found around metal poor stars. The semi-empirical mass-radius relation of small rocky exoplanets obtained by \cite{Zeng2015} shows that $1.5 R_\oplus$ corresponds to $4.2 M_\oplus$. In our sample, no planets smaller than 4.2 {\Mearth} at less than 15 days are found around stars more metallic than the Sun, which is therefore in agreement with \cite{Dawson2015}.\\
\indent Finally, one additional remark can be made. We know at least 18 planets more massive than 40 {\Mearth} with [Fe/H] < -0.4 (source : exoplanets.org), when no Neptunes are found in that metallicity range, cf Section \ref{freq}. Moreover the orbital periods of these planets are very different, from 2.96 days (WASP-98 b) to 956 days (HD 181720 b). There is therefore a turnover in the distribution that coincides with the planet desert that separates Neptunian and Jovian planets, between 40 and 60 {\Mearth}. This implies that the formation mechanism of giant planets is significantly different to that of low mass planets and reinforces the role of metallicity in planetary formation. \\

\section{Conclusion}

\indent We compiled all the known low-mass planets ($Msin(i) < 40\Mearth$) as of december 2015 with a good precision on both minimal mass (<20 \%) and host star metallicity (<0.2 dex). We studied the resulting 157 objects in the mass-metallicity diagram. The maximum mass at any given metallicity bin increases with metallicity in a seemingly linear manner (with a possible plateau below -0.5 dex), with a desert for Neptune like planets with low host-star metallicity. Observational or selection biases could not reproduce this feature, which is therefore physical in nature, as they would have either no impact or favor the detection of more massive planets that would have been located in the observed desert. \\
\indent We demonstrated that there is a dependency between this upper boundary and the planets period. The boundary shrinks at shorter periods and is more expanded at longer periods. This is in agreement with an in-situ core-accretion formation mechanism where the irradiation depleted more efficiently the protoplanetary disk of its volatile content. Alternatively, it could mean that planets around metal-poor stars form at longer periods or migrate not as quickly, and would orbit outside the period range probed by current surveys. However, a number of parameters were not considered in this study, such as planet multiplicity, planetary migrations and system architecture. \\
\indent This period dependent boundary has implications on the low mass planet frequency. The distribution in metallicity of Neptunes hosts (between 10 and 40 {\Mearth}) indicates that their frequency is likely to increase with metallicity, and that this effect is stronger for high irradiation planets. On the contrary, no such effect is visible for the Super-Earth population (<10 {\Mearth}). These results, that are corroborated by recent papers on the Kepler sample \citep{Buchhave2014,Dawson2015,Wang2015}, might change the claim that the occurence rate of Neptunes is not correlated to metallicity, as opposed to giant planets. The discrepancy with previous studies \citep{Sousa2008, Ghezzi2010, Mayor2011, Sousa2011} can be explained by a larger sample and a distinction between Neptunes and Super-Earths, that exhibit different behaviors. \\
\indent The statistical properties of low mass planets are crucial to understand planetary formation. New properties such as this exclusion zone should be now explained in the framework of planet formation models. It is of prime importance to significantly increase the sample of planets in this small mass domain with accurate parameters in order to refine and assess the robustness of the current result. Our result have important implication regarding the expected results of radial velocity surveys targeting metal-poor stars. More specifically, programs dedicated to the observation of metal-poor stars on long timescales could confirm if Neptune-like planets do exist at longer periods. Upcoming programs dedicated to transit search around bright stars (TESS, CHEOPS, NGTS) alongside ground RV facilities will enable the study of thousands of low mass planets. Not only a much larger sample but additional parameters such as stellar age, multiplicity, true mass, radius or density will provide new and crucial insights on this exclusion zone, the impact of irradiation, the differences between Neptunes and Super-Earth populations and planetary formations processes.

\section*{Acknowledgements}
     We sincerely thank Christian Marinoni for useful comments and conversations. We also gratefully acknowledge the Programme National de Plan{\'e}tologie (financial support) of CNRS/INSU.




\bibliographystyle{mnras}
\bibliography{Femass_bib} 

\begin{thebibliography}{}
\makeatletter
\relax
\def\mn@urlcharsother{\let\do\@makeother \do\$\do\&\do\#\do\^\do\_\do\%\do\~}
\def\mn@doi{\begingroup\mn@urlcharsother \@ifnextchar [ {\mn@doi@}
  {\mn@doi@[]}}
\def\mn@doi@[#1]#2{\def\@tempa{#1}\ifx\@tempa\@empty \href
  {http://dx.doi.org/#2} {doi:#2}\else \href {http://dx.doi.org/#2} {#1}\fi
  \endgroup}
\def\mn@eprint#1#2{\mn@eprint@#1:#2::\@nil}
\def\mn@eprint@arXiv#1{\href {http://arxiv.org/abs/#1} {{\tt arXiv:#1}}}
\def\mn@eprint@dblp#1{\href {http://dblp.uni-trier.de/rec/bibtex/#1.xml}
  {dblp:#1}}
\def\mn@eprint@#1:#2:#3:#4\@nil{\def\@tempa {#1}\def\@tempb {#2}\def\@tempc
  {#3}\ifx \@tempc \@empty \let \@tempc \@tempb \let \@tempb \@tempa \fi \ifx
  \@tempb \@empty \def\@tempb {arXiv}\fi \@ifundefined
  {mn@eprint@\@tempb}{\@tempb:\@tempc}{\expandafter \expandafter \csname
  mn@eprint@\@tempb\endcsname \expandafter{\@tempc}}}

\bibitem[\protect\citeauthoryear{{Adibekyan} et~al.,}{{Adibekyan}
  et~al.}{2013}]{Adibekyan2013}
{Adibekyan} V.~Z.,  et~al., 2013, \mn@doi [\aap] {10.1051/0004-6361/201322551},
  \href {http://adsabs.harvard.edu/abs/2013A%26A...560A..51A} {560, A51}

\bibitem[\protect\citeauthoryear{{Anglada-Escud{\'e}} \&
  {Tuomi}}{{Anglada-Escud{\'e}} \& {Tuomi}}{2012}]{AngladaTuomi2012}
{Anglada-Escud{\'e}} G.,  {Tuomi} M.,  2012, \mn@doi [\aap]
  {10.1051/0004-6361/201219910}, \href
  {http://adsabs.harvard.edu/abs/2012A%26A...548A..58A} {548, A58}

\bibitem[\protect\citeauthoryear{{Anglada-Escud{\'e}}
  et~al.,}{{Anglada-Escud{\'e}} et~al.}{2013}]{Anglada2013}
{Anglada-Escud{\'e}} G.,  et~al., 2013, \mn@doi [\aap]
  {10.1051/0004-6361/201321331}, \href
  {http://adsabs.harvard.edu/abs/2013A%26A...556A.126A} {556, A126}

\bibitem[\protect\citeauthoryear{{Anglada-Escud{\'e}}
  et~al.,}{{Anglada-Escud{\'e}} et~al.}{2014}]{Anglada2014}
{Anglada-Escud{\'e}} G.,  et~al., 2014, \mn@doi [\mnras]
  {10.1093/mnrasl/slu076}, \href
  {http://adsabs.harvard.edu/abs/2014MNRAS.443L..89A} {443, L89}

\bibitem[\protect\citeauthoryear{{Astudillo-Defru} et~al.,}{{Astudillo-Defru}
  et~al.}{2015}]{Astudillo2015}
{Astudillo-Defru} N.,  et~al., 2015, \mn@doi [\aap]
  {10.1051/0004-6361/201424253}, \href
  {http://adsabs.harvard.edu/abs/2015A%26A...575A.119A} {575, A119}

\bibitem[\protect\citeauthoryear{{Baglin}}{{Baglin}}{2003}]{Baglin2003}
{Baglin} A.,  2003, \mn@doi [Advances in Space Research]
  {10.1016/S0273-1177(02)00624-5}, \href
  {http://adsabs.harvard.edu/abs/2003AdSpR..31..345B} {31, 345}

\bibitem[\protect\citeauthoryear{{Bakos} et~al.,}{{Bakos}
  et~al.}{2010}]{Bakos2010}
{Bakos} G.~{\'A}.,  et~al., 2010, \mn@doi [\apj]
  {10.1088/0004-637X/710/2/1724}, \href
  {http://adsabs.harvard.edu/abs/2010ApJ...710.1724B} {710, 1724}

\bibitem[\protect\citeauthoryear{{Bakos} et~al.,}{{Bakos}
  et~al.}{2015}]{Bakos2015}
{Bakos} G.~{\'A}.,  et~al., 2015, preprint, \href
  {http://adsabs.harvard.edu/abs/2015arXiv150701024B} {} (\mn@eprint {arXiv}
  {1507.01024})

\bibitem[\protect\citeauthoryear{{Bonfils} et~al.,}{{Bonfils}
  et~al.}{2007}]{Bonfils2007}
{Bonfils} X.,  et~al., 2007, \mn@doi [\aap] {10.1051/0004-6361:20077068}, \href
  {http://adsabs.harvard.edu/abs/2007A%26A...474..293B} {474, 293}

\bibitem[\protect\citeauthoryear{{Bonfils} et~al.,}{{Bonfils}
  et~al.}{2011}]{Bonfils2011}
{Bonfils} X.,  et~al., 2011, \mn@doi [\aap] {10.1051/0004-6361/201015981},
  \href {http://adsabs.harvard.edu/abs/2011A%26A...528A.111B} {528, A111}

\bibitem[\protect\citeauthoryear{{Bonfils} et~al.,}{{Bonfils}
  et~al.}{2013}]{Bonfils2013}
{Bonfils} X.,  et~al., 2013, \mn@doi [\aap] {10.1051/0004-6361/201220237},
  \href {http://adsabs.harvard.edu/abs/2013A%26A...556A.110B} {556, A110}

\bibitem[\protect\citeauthoryear{{Borucki} \& {Koch}}{{Borucki} \&
  {Koch}}{2011}]{Borucki2011}
{Borucki} W.~J.,  {Koch} D.~G.,  2011, in {Sozzetti} A.,  {Lattanzi} M.~G.,
  {Boss} A.~P.,  eds,  IAU Symposium Vol. 276, IAU Symposium. pp 34--43,
  \mn@doi{10.1017/S1743921311019909}

\bibitem[\protect\citeauthoryear{{Borucki} et~al.,}{{Borucki}
  et~al.}{2010}]{Borucki2010}
{Borucki} W.~J.,  et~al., 2010, \mn@doi [\apjl] {10.1088/2041-8205/713/2/L126},
  \href {http://adsabs.harvard.edu/abs/2010ApJ...713L.126B} {713, L126}

\bibitem[\protect\citeauthoryear{{Bouchy} et~al.,}{{Bouchy}
  et~al.}{2009}]{Bouchy2009}
{Bouchy} F.,  et~al., 2009, \mn@doi [\aap] {10.1051/0004-6361:200810669}, \href
  {http://adsabs.harvard.edu/abs/2009A%26A...496..527B} {496, 527}

\bibitem[\protect\citeauthoryear{{Buchhave} et~al.,}{{Buchhave}
  et~al.}{2014}]{Buchhave2014}
{Buchhave} L.~A.,  et~al., 2014, \mn@doi [\nat] {10.1038/nature13254}, \href
  {http://adsabs.harvard.edu/abs/2014Natur.509..593B} {509, 593}

\bibitem[\protect\citeauthoryear{{Burt}, {Vogt}, {Butler}, {Hanson},
  {Meschiari}, {Rivera}, {Henry}  \& {Laughlin}}{{Burt}
  et~al.}{2014}]{Burt2014}
{Burt} J.,  {Vogt} S.~S.,  {Butler} R.~P.,  {Hanson} R.,  {Meschiari} S.,
  {Rivera} E.~J.,  {Henry} G.~W.,   {Laughlin} G.,  2014, \mn@doi [\apj]
  {10.1088/0004-637X/789/2/114}, \href
  {http://adsabs.harvard.edu/abs/2014ApJ...789..114B} {789, 114}

\bibitem[\protect\citeauthoryear{{Butler} et~al.,}{{Butler}
  et~al.}{2006}]{Butler2006}
{Butler} R.~P.,  et~al., 2006, \mn@doi [\apj] {10.1086/504701}, \href
  {http://adsabs.harvard.edu/abs/2006ApJ...646..505B} {646, 505}

\bibitem[\protect\citeauthoryear{{Carter}, {Winn}, {Holman}, {Fabrycky},
  {Berta}, {Burke}  \& {Nutzman}}{{Carter} et~al.}{2011}]{Carter2011}
{Carter} J.~A.,  {Winn} J.~N.,  {Holman} M.~J.,  {Fabrycky} D.,  {Berta} Z.~K.,
   {Burke} C.~J.,   {Nutzman} P.,  2011, \mn@doi [\apj]
  {10.1088/0004-637X/730/2/82}, \href
  {http://adsabs.harvard.edu/abs/2011ApJ...730...82C} {730, 82}

\bibitem[\protect\citeauthoryear{{Carter} et~al.,}{{Carter}
  et~al.}{2012}]{Carter2012}
{Carter} J.~A.,  et~al., 2012, \mn@doi [Science] {10.1126/science.1223269},
  \href {http://adsabs.harvard.edu/abs/2012Sci...337..556C} {337, 556}

\bibitem[\protect\citeauthoryear{{Courcol} et~al.,}{{Courcol}
  et~al.}{2015}]{Courcol2015}
{Courcol} B.,  et~al., 2015, \mn@doi [\aap] {10.1051/0004-6361/201526329},
  \href {http://adsabs.harvard.edu/abs/2015A%26A...581A..38C} {581, A38}

\bibitem[\protect\citeauthoryear{{Dawson}, {Chiang}  \& {Lee}}{{Dawson}
  et~al.}{2015}]{Dawson2015}
{Dawson} R.~I.,  {Chiang} E.,   {Lee} E.~J.,  2015, \mn@doi [\mnras]
  {10.1093/mnras/stv1639}, \href
  {http://adsabs.harvard.edu/abs/2015MNRAS.453.1471D} {453, 1471}

\bibitem[\protect\citeauthoryear{{Delfosse} et~al.,}{{Delfosse}
  et~al.}{2013}]{Delfosse2013}
{Delfosse} X.,  et~al., 2013, \mn@doi [\aap] {10.1051/0004-6361/201219013},
  \href {http://adsabs.harvard.edu/abs/2013A%26A...553A...8D} {553, A8}

\bibitem[\protect\citeauthoryear{{Demory} et~al.,}{{Demory}
  et~al.}{2013}]{Demory2013}
{Demory} B.-O.,  et~al., 2013, \mn@doi [\apj] {10.1088/0004-637X/768/2/154},
  \href {http://adsabs.harvard.edu/abs/2013ApJ...768..154D} {768, 154}

\bibitem[\protect\citeauthoryear{{D{\'{\i}}az} et~al.,}{{D{\'{\i}}az}
  et~al.}{2016}]{Diaz2015}
{D{\'{\i}}az} R.~F.,  et~al., 2016, \mn@doi [\aap]
  {10.1051/0004-6361/201526729}, \href
  {http://adsabs.harvard.edu/abs/2016A%26A...585A.134D} {585, A134}

\bibitem[\protect\citeauthoryear{{Dressing} et~al.,}{{Dressing}
  et~al.}{2015}]{Dressing2015}
{Dressing} C.~D.,  et~al., 2015, \mn@doi [\apj] {10.1088/0004-637X/800/2/135},
  \href {http://adsabs.harvard.edu/abs/2015ApJ...800..135D} {800, 135}

\bibitem[\protect\citeauthoryear{{Dumusque} et~al.,}{{Dumusque}
  et~al.}{2012}]{Dumusque2012}
{Dumusque} X.,  et~al., 2012, \mn@doi [\nat] {10.1038/nature11572}, \href
  {http://adsabs.harvard.edu/abs/2012Natur.491..207D} {491, 207}

\bibitem[\protect\citeauthoryear{{Dumusque} et~al.,}{{Dumusque}
  et~al.}{2014}]{Dumusque2014}
{Dumusque} X.,  et~al., 2014, \mn@doi [\apj] {10.1088/0004-637X/789/2/154},
  \href {http://adsabs.harvard.edu/abs/2014ApJ...789..154D} {789, 154}

\bibitem[\protect\citeauthoryear{{Endl} et~al.,}{{Endl}
  et~al.}{2012}]{Endl2012}
{Endl} M.,  et~al., 2012, \mn@doi [\apj] {10.1088/0004-637X/759/1/19}, \href
  {http://adsabs.harvard.edu/abs/2012ApJ...759...19E} {759, 19}

\bibitem[\protect\citeauthoryear{{Fischer} et~al.,}{{Fischer}
  et~al.}{2012}]{Fischer2012}
{Fischer} D.~A.,  et~al., 2012, \mn@doi [\apj] {10.1088/0004-637X/745/1/21},
  \href {http://adsabs.harvard.edu/abs/2012ApJ...745...21F} {745, 21}

\bibitem[\protect\citeauthoryear{{Forveille} et~al.,}{{Forveille}
  et~al.}{2009}]{Forveille2009}
{Forveille} T.,  et~al., 2009, \mn@doi [\aap] {10.1051/0004-6361:200810557},
  \href {http://adsabs.harvard.edu/abs/2009A%26A...493..645F} {493, 645}

\bibitem[\protect\citeauthoryear{{Forveille} et~al.,}{{Forveille}
  et~al.}{2011}]{Forveille2011}
{Forveille} T.,  et~al., 2011, preprint, \href
  {http://adsabs.harvard.edu/abs/2011arXiv1109.2505F} {} (\mn@eprint {arXiv}
  {1109.2505})

\bibitem[\protect\citeauthoryear{{Fulton} et~al.,}{{Fulton}
  et~al.}{2015}]{Fulton2015}
{Fulton} B.~J.,  et~al., 2015, \mn@doi [\apj] {10.1088/0004-637X/805/2/175},
  \href {http://adsabs.harvard.edu/abs/2015ApJ...805..175F} {805, 175}

\bibitem[\protect\citeauthoryear{{Ghezzi}, {Cunha}, {Smith}, {de Ara{\'u}jo},
  {Schuler}  \& {de la Reza}}{{Ghezzi} et~al.}{2010}]{Ghezzi2010}
{Ghezzi} L.,  {Cunha} K.,  {Smith} V.~V.,  {de Ara{\'u}jo} F.~X.,  {Schuler}
  S.~C.,   {de la Reza} R.,  2010, \mn@doi [\apj]
  {10.1088/0004-637X/720/2/1290}, \href
  {http://adsabs.harvard.edu/abs/2010ApJ...720.1290G} {720, 1290}

\bibitem[\protect\citeauthoryear{{Gonzalez}}{{Gonzalez}}{1997}]{Gonzalez1997}
{Gonzalez} G.,  1997, \mnras, \href
  {http://adsabs.harvard.edu/abs/1997MNRAS.285..403G} {285, 403}

\bibitem[\protect\citeauthoryear{{Hadden} \& {Lithwick}}{{Hadden} \&
  {Lithwick}}{2014}]{HaddenLithwick2014}
{Hadden} S.,  {Lithwick} Y.,  2014, \mn@doi [\apj]
  {10.1088/0004-637X/787/1/80}, \href
  {http://adsabs.harvard.edu/abs/2014ApJ...787...80H} {787, 80}

\bibitem[\protect\citeauthoryear{{Hartman} et~al.,}{{Hartman}
  et~al.}{2011}]{Hartman2011}
{Hartman} J.~D.,  et~al., 2011, \mn@doi [\apj] {10.1088/0004-637X/728/2/138},
  \href {http://adsabs.harvard.edu/abs/2011ApJ...728..138H} {728, 138}

\bibitem[\protect\citeauthoryear{{Haywood} et~al.,}{{Haywood}
  et~al.}{2014}]{Haywood2014}
{Haywood} R.~D.,  et~al., 2014, \mn@doi [\mnras] {10.1093/mnras/stu1320}, \href
  {http://adsabs.harvard.edu/abs/2014MNRAS.443.2517H} {443, 2517}

\bibitem[\protect\citeauthoryear{{H{\'e}brard} et~al.,}{{H{\'e}brard}
  et~al.}{2010}]{Hebrard2010}
{H{\'e}brard} G.,  et~al., 2010, \mn@doi [\aap] {10.1051/0004-6361/200913525},
  \href {http://adsabs.harvard.edu/abs/2010A%26A...512A..46H} {512, A46}

\bibitem[\protect\citeauthoryear{{Hirano} et~al.,}{{Hirano}
  et~al.}{2012}]{Hirano2012}
{Hirano} T.,  et~al., 2012, \mn@doi [\apjl] {10.1088/2041-8205/759/2/L36},
  \href {http://adsabs.harvard.edu/abs/2012ApJ...759L..36H} {759, L36}

\bibitem[\protect\citeauthoryear{{Howard} et~al.,}{{Howard}
  et~al.}{2009}]{Howard2009}
{Howard} A.~W.,  et~al., 2009, \mn@doi [\apj] {10.1088/0004-637X/696/1/75},
  \href {http://adsabs.harvard.edu/abs/2009ApJ...696...75H} {696, 75}

\bibitem[\protect\citeauthoryear{{Howard} et~al.,}{{Howard}
  et~al.}{2011a}]{Howard2011a}
{Howard} A.~W.,  et~al., 2011a, \mn@doi [\apj] {10.1088/0004-637X/726/2/73},
  \href {http://adsabs.harvard.edu/abs/2011ApJ...726...73H} {726, 73}

\bibitem[\protect\citeauthoryear{{Howard} et~al.,}{{Howard}
  et~al.}{2011b}]{Howard2011b}
{Howard} A.~W.,  et~al., 2011b, \mn@doi [\apj] {10.1088/0004-637X/730/1/10},
  \href {http://adsabs.harvard.edu/abs/2011ApJ...730...10H} {730, 10}

\bibitem[\protect\citeauthoryear{{Howard} et~al.,}{{Howard}
  et~al.}{2014}]{Howard2014}
{Howard} A.~W.,  et~al., 2014, \mn@doi [\apj] {10.1088/0004-637X/794/1/51},
  \href {http://adsabs.harvard.edu/abs/2014ApJ...794...51H} {794, 51}

\bibitem[\protect\citeauthoryear{{Huber} et~al.,}{{Huber}
  et~al.}{2013}]{Huber2013}
{Huber} D.,  et~al., 2013, \mn@doi [Science] {10.1126/science.1242066}, \href
  {http://adsabs.harvard.edu/abs/2013Sci...342..331H} {342, 331}

\bibitem[\protect\citeauthoryear{{Jenkins} et~al.,}{{Jenkins}
  et~al.}{2013}]{Jenkins2013}
{Jenkins} J.~S.,  et~al., 2013, \mn@doi [\apj] {10.1088/0004-637X/766/2/67},
  \href {http://adsabs.harvard.edu/abs/2013ApJ...766...67J} {766, 67}

\bibitem[\protect\citeauthoryear{{Johnson} \& {Apps}}{{Johnson} \&
  {Apps}}{2009}]{Johnson2009}
{Johnson} J.~A.,  {Apps} K.,  2009, \mn@doi [\apj]
  {10.1088/0004-637X/699/2/933}, \href
  {http://adsabs.harvard.edu/abs/2009ApJ...699..933J} {699, 933}

\bibitem[\protect\citeauthoryear{{Laws}, {Gonzalez}, {Walker}, {Tyagi},
  {Dodsworth}, {Snider}  \& {Suntzeff}}{{Laws} et~al.}{2003}]{Laws2003}
{Laws} C.,  {Gonzalez} G.,  {Walker} K.~M.,  {Tyagi} S.,  {Dodsworth} J.,
  {Snider} K.,   {Suntzeff} N.~B.,  2003, \mn@doi [\aj] {10.1086/374626}, \href
  {http://adsabs.harvard.edu/abs/2003AJ....125.2664L} {125, 2664}

\bibitem[\protect\citeauthoryear{{Lissauer} et~al.,}{{Lissauer}
  et~al.}{2013}]{Lissauer2013}
{Lissauer} J.~J.,  et~al., 2013, \mn@doi [\apj] {10.1088/0004-637X/770/2/131},
  \href {http://adsabs.harvard.edu/abs/2013ApJ...770..131L} {770, 131}

\bibitem[\protect\citeauthoryear{{Lo Curto} et~al.,}{{Lo Curto}
  et~al.}{2010}]{LoCurto2010}
{Lo Curto} G.,  et~al., 2010, \mn@doi [\aap] {10.1051/0004-6361/200913523},
  \href {http://adsabs.harvard.edu/abs/2010A%26A...512A..48L} {512, A48}

\bibitem[\protect\citeauthoryear{{Lo Curto} et~al.,}{{Lo Curto}
  et~al.}{2013}]{LoCurto2013}
{Lo Curto} G.,  et~al., 2013, \mn@doi [\aap] {10.1051/0004-6361/201220415},
  \href {http://adsabs.harvard.edu/abs/2013A%26A...551A..59L} {551, A59}

\bibitem[\protect\citeauthoryear{{Lovis} et~al.,}{{Lovis}
  et~al.}{2006}]{Lovis2006}
{Lovis} C.,  et~al., 2006, \mn@doi [\nat] {10.1038/nature04828}, \href
  {http://adsabs.harvard.edu/abs/2006Natur.441..305L} {441, 305}

\bibitem[\protect\citeauthoryear{{Lovis} et~al.,}{{Lovis}
  et~al.}{2011}]{Lovis2011}
{Lovis} C.,  et~al., 2011, \mn@doi [\aap] {10.1051/0004-6361/201015577}, \href
  {http://adsabs.harvard.edu/abs/2011A%26A...528A.112L} {528, A112}

\bibitem[\protect\citeauthoryear{{Maness}, {Marcy}, {Ford}, {Hauschildt},
  {Shreve}, {Basri}, {Butler}  \& {Vogt}}{{Maness} et~al.}{2007}]{Maness2007}
{Maness} H.~L.,  {Marcy} G.~W.,  {Ford} E.~B.,  {Hauschildt} P.~H.,  {Shreve}
  A.~T.,  {Basri} G.~B.,  {Butler} R.~P.,   {Vogt} S.~S.,  2007, \mn@doi
  [\pasp] {10.1086/510689}, \href
  {http://adsabs.harvard.edu/abs/2007PASP..119...90M} {119, 90}

\bibitem[\protect\citeauthoryear{{Marcy} et~al.,}{{Marcy}
  et~al.}{2014}]{Marcy2014}
{Marcy} G.~W.,  et~al., 2014, \mn@doi [\apjs] {10.1088/0067-0049/210/2/20},
  \href {http://adsabs.harvard.edu/abs/2014ApJS..210...20M} {210, 20}

\bibitem[\protect\citeauthoryear{{Masuda}}{{Masuda}}{2014}]{Masuda2014}
{Masuda} K.,  2014, \mn@doi [\apj] {10.1088/0004-637X/783/1/53}, \href
  {http://adsabs.harvard.edu/abs/2014ApJ...783...53M} {783, 53}

\bibitem[\protect\citeauthoryear{{Masuda}, {Hirano}, {Taruya}, {Nagasawa}  \&
  {Suto}}{{Masuda} et~al.}{2013}]{Masuda2013}
{Masuda} K.,  {Hirano} T.,  {Taruya} A.,  {Nagasawa} M.,   {Suto} Y.,  2013,
  \mn@doi [\apj] {10.1088/0004-637X/778/2/185}, \href
  {http://adsabs.harvard.edu/abs/2013ApJ...778..185M} {778, 185}

\bibitem[\protect\citeauthoryear{{Mayor} et~al.,}{{Mayor}
  et~al.}{2009}]{Mayor2009}
{Mayor} M.,  et~al., 2009, \mn@doi [\aap] {10.1051/0004-6361:200810451}, \href
  {http://adsabs.harvard.edu/abs/2009A%26A...493..639M} {493, 639}

\bibitem[\protect\citeauthoryear{{Mayor} et~al.,}{{Mayor}
  et~al.}{2011}]{Mayor2011}
{Mayor} M.,  et~al., 2011, preprint, \href
  {http://adsabs.harvard.edu/abs/2011arXiv1109.2497M} {} (\mn@eprint {arXiv}
  {1109.2497})

\bibitem[\protect\citeauthoryear{{Melo} et~al.,}{{Melo}
  et~al.}{2007}]{Melo2007}
{Melo} C.,  et~al., 2007, \mn@doi [\aap] {10.1051/0004-6361:20066845}, \href
  {http://adsabs.harvard.edu/abs/2007A%26A...467..721M} {467, 721}

\bibitem[\protect\citeauthoryear{{Mordasini} et~al.,}{{Mordasini}
  et~al.}{2011}]{Mordasini2011}
{Mordasini} C.,  et~al., 2011, \mn@doi [\aap] {10.1051/0004-6361/200913521},
  \href {http://adsabs.harvard.edu/abs/2011A%26A...526A.111M} {526, A111}

\bibitem[\protect\citeauthoryear{{Mordasini}, {Alibert}, {Georgy}, {Dittkrist},
  {Klahr}  \& {Henning}}{{Mordasini} et~al.}{2012}]{Mordasini2012}
{Mordasini} C.,  {Alibert} Y.,  {Georgy} C.,  {Dittkrist} K.-M.,  {Klahr} H.,
  {Henning} T.,  2012, \mn@doi [\aap] {10.1051/0004-6361/201118464}, \href
  {http://adsabs.harvard.edu/abs/2012A%26A...547A.112M} {547, A112}

\bibitem[\protect\citeauthoryear{{Mortier} et~al.,}{{Mortier}
  et~al.}{2016}]{Mortier2016}
{Mortier} A.,  et~al., 2016, \mn@doi [\aap] {10.1051/0004-6361/201526905},
  \href {http://adsabs.harvard.edu/abs/2016A%26A...585A.135M} {585, A135}

\bibitem[\protect\citeauthoryear{{Motalebi} et~al.,}{{Motalebi}
  et~al.}{2015}]{Motalebi2015}
{Motalebi} F.,  et~al., 2015, \mn@doi [\aap] {10.1051/0004-6361/201526822},
  \href {http://adsabs.harvard.edu/abs/2015A%26A...584A..72M} {584, A72}

\bibitem[\protect\citeauthoryear{{Neves} et~al.,}{{Neves}
  et~al.}{2012}]{Neves2012}
{Neves} V.,  et~al., 2012, \mn@doi [\aap] {10.1051/0004-6361/201118115}, \href
  {http://adsabs.harvard.edu/abs/2012A%26A...538A..25N} {538, A25}

\bibitem[\protect\citeauthoryear{{O'Toole} et~al.,}{{O'Toole}
  et~al.}{2009a}]{OToole2009a}
{O'Toole} S.,  et~al., 2009a, \mn@doi [\apj] {10.1088/0004-637X/697/2/1263},
  \href {http://adsabs.harvard.edu/abs/2009ApJ...697.1263O} {697, 1263}

\bibitem[\protect\citeauthoryear{{O'Toole}, {Jones}, {Tinney}, {Butler},
  {Marcy}, {Carter}, {Bailey}  \& {Wittenmyer}}{{O'Toole}
  et~al.}{2009b}]{OToole2009b}
{O'Toole} S.~J.,  {Jones} H.~R.~A.,  {Tinney} C.~G.,  {Butler} R.~P.,  {Marcy}
  G.~W.,  {Carter} B.,  {Bailey} J.,   {Wittenmyer} R.~A.,  2009b, \mn@doi
  [\apj] {10.1088/0004-637X/701/2/1732}, \href
  {http://adsabs.harvard.edu/abs/2009ApJ...701.1732O} {701, 1732}

\bibitem[\protect\citeauthoryear{{Pepe} et~al.,}{{Pepe}
  et~al.}{2007}]{Pepe2007}
{Pepe} F.,  et~al., 2007, \mn@doi [\aap] {10.1051/0004-6361:20066194}, \href
  {http://adsabs.harvard.edu/abs/2007A%26A...462..769P} {462, 769}

\bibitem[\protect\citeauthoryear{{Pepe} et~al.,}{{Pepe}
  et~al.}{2011}]{Pepe2011}
{Pepe} F.,  et~al., 2011, \mn@doi [\aap] {10.1051/0004-6361/201117055}, \href
  {http://adsabs.harvard.edu/abs/2011A%26A...534A..58P} {534, A58}

\bibitem[\protect\citeauthoryear{{Pepe} et~al.,}{{Pepe}
  et~al.}{2013}]{Pepe2013}
{Pepe} F.,  et~al., 2013, \mn@doi [\nat] {10.1038/nature12768}, \href
  {http://adsabs.harvard.edu/abs/2013Natur.503..377P} {503, 377}

\bibitem[\protect\citeauthoryear{{Pollack}, {Hubickyj}, {Bodenheimer},
  {Lissauer}, {Podolak}  \& {Greenzweig}}{{Pollack} et~al.}{1996}]{Pollack1996}
{Pollack} J.~B.,  {Hubickyj} O.,  {Bodenheimer} P.,  {Lissauer} J.~J.,
  {Podolak} M.,   {Greenzweig} Y.,  1996, \mn@doi [\icarus]
  {10.1006/icar.1996.0190}, \href
  {http://adsabs.harvard.edu/abs/1996Icar..124...62P} {124, 62}

\bibitem[\protect\citeauthoryear{{Porto de Mello}, {da Silva}, {da Silva}  \&
  {de Nader}}{{Porto de Mello} et~al.}{2014}]{deMello2014}
{Porto de Mello} G.~F.,  {da Silva} R.,  {da Silva} L.,   {de Nader} R.~V.,
  2014, \mn@doi [\aap] {10.1051/0004-6361/201322277}, \href
  {http://adsabs.harvard.edu/abs/2014A%26A...563A..52P} {563, A52}

\bibitem[\protect\citeauthoryear{{Rivera}, {Laughlin}, {Butler}, {Vogt},
  {Haghighipour}  \& {Meschiari}}{{Rivera} et~al.}{2010}]{Rivera2010b}
{Rivera} E.~J.,  {Laughlin} G.,  {Butler} R.~P.,  {Vogt} S.~S.,  {Haghighipour}
  N.,   {Meschiari} S.,  2010, \mn@doi [\apj] {10.1088/0004-637X/719/1/890},
  \href {http://adsabs.harvard.edu/abs/2010ApJ...719..890R} {719, 890}

\bibitem[\protect\citeauthoryear{{Robertson} \& {Mahadevan}}{{Robertson} \&
  {Mahadevan}}{2014}]{Robertson2014}
{Robertson} P.,  {Mahadevan} S.,  2014, \mn@doi [\apjl]
  {10.1088/2041-8205/793/2/L24}, \href
  {http://adsabs.harvard.edu/abs/2014ApJ...793L..24R} {793, L24}

\bibitem[\protect\citeauthoryear{{Sanchis-Ojeda}, {Rappaport}, {Winn},
  {Levine}, {Kotson}, {Latham}  \& {Buchhave}}{{Sanchis-Ojeda}
  et~al.}{2013}]{Sanchis2013}
{Sanchis-Ojeda} R.,  {Rappaport} S.,  {Winn} J.~N.,  {Levine} A.,  {Kotson}
  M.~C.,  {Latham} D.~W.,   {Buchhave} L.~A.,  2013, \mn@doi [\apj]
  {10.1088/0004-637X/774/1/54}, \href
  {http://adsabs.harvard.edu/abs/2013ApJ...774...54S} {774, 54}

\bibitem[\protect\citeauthoryear{{Santos}, {Israelian}, {Mayor}, {Bento},
  {Almeida}, {Sousa}  \& {Ecuvillon}}{{Santos} et~al.}{2005}]{Santos2005}
{Santos} N.~C.,  {Israelian} G.,  {Mayor} M.,  {Bento} J.~P.,  {Almeida} P.~C.,
   {Sousa} S.~G.,   {Ecuvillon} A.,  2005, \mn@doi [\aap]
  {10.1051/0004-6361:20052895}, \href
  {http://adsabs.harvard.edu/abs/2005A%26A...437.1127S} {437, 1127}

\bibitem[\protect\citeauthoryear{{Santos} et~al.,}{{Santos}
  et~al.}{2013}]{SWEET}
{Santos} N.~C.,  et~al., 2013, \mn@doi [\aap] {10.1051/0004-6361/201321286},
  \href {http://cdsads.u-strasbg.fr/abs/2013A%26A...556A.150S} {556, A150}

\bibitem[\protect\citeauthoryear{{S{\'e}gransan} et~al.,}{{S{\'e}gransan}
  et~al.}{2011}]{Segransan2011}
{S{\'e}gransan} D.,  et~al., 2011, \mn@doi [\aap]
  {10.1051/0004-6361/200913580}, \href
  {http://adsabs.harvard.edu/abs/2011A%26A...535A..54S} {535, A54}

\bibitem[\protect\citeauthoryear{{Soubiran}, {Le Campion}, {Cayrel de Strobel}
  \& {Caillo}}{{Soubiran} et~al.}{2010}]{Soubiran2010}
{Soubiran} C.,  {Le Campion} J.-F.,  {Cayrel de Strobel} G.,   {Caillo} A.,
  2010, \mn@doi [\aap] {10.1051/0004-6361/201014247}, \href
  {http://adsabs.harvard.edu/abs/2010A%26A...515A.111S} {515, A111}

\bibitem[\protect\citeauthoryear{{Sousa} et~al.,}{{Sousa}
  et~al.}{2008}]{Sousa2008}
{Sousa} S.~G.,  et~al., 2008, \mn@doi [\aap] {10.1051/0004-6361:200809698},
  \href {http://adsabs.harvard.edu/abs/2008A%26A...487..373S} {487, 373}

\bibitem[\protect\citeauthoryear{{Sousa}, {Santos}, {Israelian}, {Mayor}  \&
  {Udry}}{{Sousa} et~al.}{2011}]{Sousa2011}
{Sousa} S.~G.,  {Santos} N.~C.,  {Israelian} G.,  {Mayor} M.,   {Udry} S.,
  2011, \mn@doi [\aap] {10.1051/0004-6361/201117699}, \href
  {http://adsabs.harvard.edu/abs/2011A%26A...533A.141S} {533, A141}

\bibitem[\protect\citeauthoryear{{Tinney}, {Butler}, {Jones}, {Wittenmyer},
  {O'Toole}, {Bailey}  \& {Carter}}{{Tinney} et~al.}{2011}]{Tinney2011}
{Tinney} C.~G.,  {Butler} R.~P.,  {Jones} H.~R.~A.,  {Wittenmyer} R.~A.,
  {O'Toole} S.,  {Bailey} J.,   {Carter} B.~D.,  2011, \mn@doi [\apj]
  {10.1088/0004-637X/727/2/103}, \href
  {http://adsabs.harvard.edu/abs/2011ApJ...727..103T} {727, 103}

\bibitem[\protect\citeauthoryear{{Tuomi}}{{Tuomi}}{2014}]{Tuomi2014}
{Tuomi} M.,  2014, \mn@doi [\mnras] {10.1093/mnrasl/slu014}, \href
  {http://adsabs.harvard.edu/abs/2014MNRAS.440L...1T} {440, L1}

\bibitem[\protect\citeauthoryear{{Tuomi}, {Anglada-Escud{\'e}}, {Gerlach},
  {Jones}, {Reiners}, {Rivera}, {Vogt}  \& {Butler}}{{Tuomi}
  et~al.}{2013}]{Tuomi2013}
{Tuomi} M.,  {Anglada-Escud{\'e}} G.,  {Gerlach} E.,  {Jones} H.~R.~A.,
  {Reiners} A.,  {Rivera} E.~J.,  {Vogt} S.~S.,   {Butler} R.~P.,  2013,
  \mn@doi [\aap] {10.1051/0004-6361/201220268}, \href
  {http://adsabs.harvard.edu/abs/2013A%26A...549A..48T} {549, A48}

\bibitem[\protect\citeauthoryear{{Valenti} et~al.,}{{Valenti}
  et~al.}{2009}]{Valenti2009}
{Valenti} J.~A.,  et~al., 2009, \mn@doi [\apj] {10.1088/0004-637X/702/2/989},
  \href {http://adsabs.harvard.edu/abs/2009ApJ...702..989V} {702, 989}

\bibitem[\protect\citeauthoryear{{Van Grootel} et~al.,}{{Van Grootel}
  et~al.}{2014}]{VanGrootel2014}
{Van Grootel} V.,  et~al., 2014, \mn@doi [\apj] {10.1088/0004-637X/786/1/2},
  \href {http://adsabs.harvard.edu/abs/2014ApJ...786....2V} {786, 2}

\bibitem[\protect\citeauthoryear{{Vanderburg} et~al.,}{{Vanderburg}
  et~al.}{2015}]{Vanderburg2015}
{Vanderburg} A.,  et~al., 2015, \mn@doi [\apj] {10.1088/0004-637X/800/1/59},
  \href {http://adsabs.harvard.edu/abs/2015ApJ...800...59V} {800, 59}

\bibitem[\protect\citeauthoryear{{Vogt} et~al.,}{{Vogt}
  et~al.}{2010}]{Vogt2010}
{Vogt} S.~S.,  et~al., 2010, \mn@doi [\apj] {10.1088/0004-637X/708/2/1366},
  \href {http://adsabs.harvard.edu/abs/2010ApJ...708.1366V} {708, 1366}

\bibitem[\protect\citeauthoryear{{Vogt} et~al.,}{{Vogt}
  et~al.}{2015}]{Vogt2015}
{Vogt} S.~S.,  et~al., 2015, \mn@doi [\apj] {10.1088/0004-637X/814/1/12}, \href
  {http://adsabs.harvard.edu/abs/2015ApJ...814...12V} {814, 12}

\bibitem[\protect\citeauthoryear{{Wang} \& {Fischer}}{{Wang} \&
  {Fischer}}{2015}]{Wang2015}
{Wang} J.,  {Fischer} D.~A.,  2015, \mn@doi [\aj] {10.1088/0004-6256/149/1/14},
  \href {http://adsabs.harvard.edu/abs/2015AJ....149...14W} {149, 14}

\bibitem[\protect\citeauthoryear{{Wittenmyer} et~al.,}{{Wittenmyer}
  et~al.}{2014}]{Wittenmyer2014}
{Wittenmyer} R.~A.,  et~al., 2014, \mn@doi [\apj]
  {10.1088/0004-637X/791/2/114}, \href
  {http://adsabs.harvard.edu/abs/2014ApJ...791..114W} {791, 114}

\bibitem[\protect\citeauthoryear{{Wright}, {Upadhyay}, {Marcy}, {Fischer},
  {Ford}  \& {Johnson}}{{Wright} et~al.}{2009}]{Wright2009}
{Wright} J.~T.,  {Upadhyay} S.,  {Marcy} G.~W.,  {Fischer} D.~A.,  {Ford}
  E.~B.,   {Johnson} J.~A.,  2009, \mn@doi [\apj]
  {10.1088/0004-637X/693/2/1084}, \href
  {http://adsabs.harvard.edu/abs/2009ApJ...693.1084W} {693, 1084}

\bibitem[\protect\citeauthoryear{{Xie}}{{Xie}}{2014}]{Xie2014}
{Xie} J.-W.,  2014, \mn@doi [\apjs] {10.1088/0067-0049/210/2/25}, \href
  {http://adsabs.harvard.edu/abs/2014ApJS..210...25X} {210, 25}

\bibitem[\protect\citeauthoryear{{Zeng}, {Sasselov}  \& {Jacobsen}}{{Zeng}
  et~al.}{2015}]{Zeng2015}
{Zeng} L.,  {Sasselov} D.,   {Jacobsen} S.,  2015, preprint, \href
  {http://adsabs.harvard.edu/abs/2015arXiv151208827Z} {} (\mn@eprint {arXiv}
  {1512.08827})

\makeatother
\end{thebibliography}


\appendix

\onecolumn

\begin{table}
\centering
\caption{Annex : Parameters of the sample}
\begin{tabular}{lcccc}
\hline
Name & Mass [\Mearth] & [Fe/H] & Period [days] & References \\
\hline
Kepler-78 b & $ 1.86 \pm 0.30 $ & $ -0.14 \pm 0.08 $ & $0.35$ & \citet{Pepe2013}, \citet{Sanchis2013} \\
55 Cnc e & $ 8.32 \pm 0.39 $ & $ 0.33 \pm 0.07 $ & $0.74$ & \citet{Endl2012}, \citet{SWEET} \\
Kepler-10 b & $ 3.33 \pm 0.49 $ & $ -0.15 \pm 0.04 $ & $0.84$ & \citet{Dumusque2014}, \citet{SWEET} \\
GJ 1214 b & $ 6.47 \pm 1.00 $ & $ 0.01 \pm 0.20 $ & $1.58$ & \citet{Carter2011}, \citet{SWEET} \\
GJ 876 d & $ 5.85 \pm 0.39 $ & $ 0.15 \pm 0.10 $ & $1.94$ & \citet{Rivera2010b}, \citet{SWEET} \\
GJ 436 b & $ 23.06 \pm 1.01 $ & $ 0.01 \pm 0.20 $ & $2.64$ & \citet{Maness2007}, \citet{SWEET} \\
GJ 3634 b & $ 7.05 \pm 0.87 $ & $ -0.04 \pm 0.20 $ & $2.65$ & \citet{Bonfils2011}, \citet{SWEET} \\
GJ 581 e & $ 1.95 \pm 0.22 $ & $ 0.21 \pm 0.10 $ & $3.15$ & \citet{Forveille2011}, \citet{SWEET} \\
HATS-7 b & $ 38.00 \pm 3.80 $ & $ 0.25 \pm 0.08 $ & $3.18$ & \citet{Bakos2015} \\
Kepler-4 b & $ 24.50 \pm 3.80 $ & $ 0.17 \pm 0.06 $ & $3.20$ & \citet{Borucki2010}, \citet{SWEET} \\
alpha Cen B b & $ 1.13 \pm 0.10 $ & $ 0.16 \pm 0.04 $ & $3.24$ & \citet{Dumusque2012}, \citet{SWEET} \\
61 Vir b & $ 5.10 \pm 0.60 $ & $ 0.01 \pm 0.05 $ & $4.20$ & \citet{Vogt2010}, \citet{SWEET} \\
61 Vir c & $ 18.20 \pm 1.10 $ & $ 0.01 \pm 0.05 $ & $38.00$ & \citet{Vogt2010}, \citet{SWEET} \\
61 Vir d & $ 22.90 \pm 2.60 $ & $ 0.01 \pm 0.05 $ & $123.00$ & \citet{Vogt2010}, \citet{SWEET} \\
BD -08 2823 b & $ 14.60 \pm 1.01 $ & $ 0.00 \pm 0.08 $ & $5.60$ & \citet{Hebrard2010}, \citet{SWEET} \\
BD-061339 b & $ 6.30 \pm 0.80 $ & $ -0.14 \pm 0.17 $ & $3.87$ & \citet{Tuomi2014}, \citet{SWEET} \\
CoRoT-7 c & $ 13.56 \pm 1.08 $ & $ 0.02 \pm 0.02 $ & $3.70$ & \citet{Haywood2014}, \citet{SWEET} \\
GJ 15 A b & $ 5.34 \pm 0.76 $ & $ -0.32 \pm 0.17 $ & $11.44$ & \citet{Howard2014} \\
GJ 160.2 b & $ 10.20 \pm 2.00 $ & $ 0.00 \pm 0.15 $ & $5.24$ & \citet{Tuomi2014}, \citet{Soubiran2010} \\
GJ 163 b & $ 10.77 \pm 0.85 $ & $ -0.02 \pm 0.20 $ & $8.63$ & \citet{Bonfils2013}, \citet{SWEET} \\
GJ 163 c & $ 6.85 \pm 0.99 $ & $ -0.02 \pm 0.20 $ & $25.63$ & \citet{Bonfils2013}, \citet{SWEET} \\
GJ 163 d & $ 29.43 \pm 4.05 $ & $ -0.02 \pm 0.20 $ & $603.95$ & \citet{Bonfils2013}, \citet{SWEET} \\
GJ 176 b & $ 8.40 \pm 1.00 $ & $ -0.01 \pm 0.10 $ & $8.80$ & \citet{Forveille2009}, \citet{SWEET} \\
GJ 3293 b & $ 24.00 \pm 1.70 $ & $ 0.02 \pm 0.09 $ & $30.60$ & \citet{Astudillo2015} \\
GJ 3293 d & $ 22.30 \pm 1.70 $ & $ 0.02 \pm 0.09 $ & $124.00$ & \citet{Astudillo2015} \\
GJ 3470 b & $ 13.90 \pm 1.50 $ & $ 0.08 \pm 0.10 $ & $3.30$ & \citet{Demory2013}, \citet{SWEET} \\
GJ 433 b & $ 5.78 \pm 0.47 $ & $ -0.17 \pm 0.10 $ & $7.37$ & \citet{Delfosse2013}, \citet{SWEET} \\
GJ 581 b & $ 15.86 \pm 0.72 $ & $ 0.21 \pm 0.10 $ & $5.37$ & \citet{Forveille2011}, \citet{SWEET} \\
GJ 581 c & $ 5.33 \pm 0.38 $ & $ 0.21 \pm 0.10 $ & $12.92$ & \citet{Forveille2011}, \citet{SWEET} \\
GJ 667 C b & $ 5.56 \pm 0.34 $ & $ -0.53 \pm 0.10 $ & $7.20$ & \citet{Robertson2014}, \citet{SWEET} \\
GJ 667 C c & $ 4.15 \pm 0.68 $ & $ -0.53 \pm 0.10 $ & $28.10$ & \citet{Robertson2014}, \citet{SWEET} \\
GJ 667C d & $ 5.10 \pm 0.60 $ & $ -0.53 \pm 0.10 $ & $91.61$ & \citet{Anglada2013}, \citet{SWEET} \\
GJ 667C e & $ 2.70 \pm 0.50 $ & $ -0.53 \pm 0.10 $ & $62.24$ & \citet{Anglada2013}, \citet{SWEET} \\
GJ 667C f & $ 3.80 \pm 0.40 $ & $ -0.53 \pm 0.10 $ & $28.14$ & \citet{Anglada2013}, \citet{SWEET} \\
GJ 667C g & $ 4.60 \pm 0.80 $ & $ -0.53 \pm 0.10 $ & $256.20$ & \citet{Anglada2013}, \citet{SWEET} \\
GJ 674 b & $ 11.09 \pm 0.24 $ & $ -0.25 \pm 0.10 $ & $4.70$ & \citet{Bonfils2007}, \citet{SWEET} \\
GJ 676A d & $ 4.40 \pm 0.70 $ & $ 0.08 \pm 0.20 $ & $3.60$ & \citet{AngladaTuomi2012}, \citet{SWEET} \\
GJ 676A e & $ 11.50 \pm 1.50 $ & $ 0.08 \pm 0.20 $ & $35.37$ & \citet{AngladaTuomi2012}, \citet{SWEET} \\
GJ 832 c & $ 5.00 \pm 1.00 $ & $ -0.19 \pm 0.10 $ & $35.70$ & \citet{Wittenmyer2014}, \citet{SWEET} \\
GJ 876 e & $ 12.47 \pm 1.62 $ & $ 0.15 \pm 0.10 $ & $124.26$ & \citet{Rivera2010b}, \citet{SWEET} \\
Gl 687 b & $ 18.00 \pm 2.00 $ & $ -0.09 \pm 0.15 $ & $38.10$ & \citet{Burt2014} \\
Gl 785 b & $ 21.60 \pm 2.00 $ & $ 0.08 \pm 0.03 $ & $74.40$ & \citet{Howard2011b} \\
HAT-P-11 b & $ 26.22 \pm 2.86 $ & $ 0.26 \pm 0.08 $ & $4.89$ & \citet{Bakos2010}, \citet{SWEET} \\
HAT-P-26 b & $ 18.70 \pm 2.20 $ & $ 0.01 \pm 0.04 $ & $4.23$ & \citet{Hartman2011}, \citet{SWEET} \\
HD 10180 c & $ 13.19 \pm 0.62 $ & $ 0.08 \pm 0.01 $ & $5.76$ & \citet{Lovis2011}, \citet{SWEET} \\
HD 10180 d & $ 11.97 \pm 0.77 $ & $ 0.08 \pm 0.01 $ & $16.36$ & \citet{Lovis2011}, \citet{SWEET} \\
HD 10180 e & $ 25.36 \pm 1.37 $ & $ 0.08 \pm 0.01 $ & $49.75$ & \citet{Lovis2011}, \citet{SWEET} \\
HD 10180 f & $ 23.62 \pm 1.66 $ & $ 0.08 \pm 0.01 $ & $122.72$ & \citet{Lovis2011}, \citet{SWEET} \\
HD 10180 g & $ 21.41 \pm 2.97 $ & $ 0.08 \pm 0.01 $ & $602.00$ & \citet{Lovis2011}, \citet{SWEET} \\
HD 102365 b & $ 16.20 \pm 2.58 $ & $ -0.29 \pm 0.02 $ & $122.10$ & \citet{Tinney2011}, \citet{SWEET} \\
HD 103197 b & $ 31.22 \pm 1.90 $ & $ 0.22 \pm 0.04 $ & $47.84$ & \citet{Mordasini2011}, \citet{SWEET} \\
HD 109271 b & $ 17.00 \pm 1.00 $ & $ 0.10 \pm 0.01 $ & $7.90$ & \citet{LoCurto2013}, \citet{SWEET} \\
HD 109271 c & $ 24.00 \pm 2.00 $ & $ 0.10 \pm 0.01 $ & $30.90$ & \citet{LoCurto2013}, \citet{SWEET} \\
HD 11964 c & $ 24.49 \pm 3.51 $ & $ 0.14 \pm 0.05 $ & $37.91$ & \citet{Wright2009}, \citet{SWEET} \\
HD 125595 b & $ 13.25 \pm 1.37 $ & $ 0.10 \pm 0.14 $ & $9.67$ & \citet{Segransan2011}, \citet{SWEET} \\
HD 125612 c & $ 18.45 \pm 3.28 $ & $ 0.24 \pm 0.01 $ & $4.15$ & \citet{LoCurto2010}, \citet{SWEET} \\
\hline
\end{tabular}
\end{table}

\newpage
\begin{table}
\centering
\contcaption{Annex : Parameters of the sample}
\begin{tabular}{lcccc}
\hline
Name & Mass [\Mearth] & [Fe/H] & Period [days] & References \\
\hline
HD 134060 b & $ 11.17 \pm 0.66 $ & $ 0.14 \pm 0.01 $ & $38.00$ & \citet{Mayor2011}, \citet{SWEET} \\
HD 134606 b & $ 2.37 \pm 0.28 $ & $ 0.27 \pm 0.02 $ & $4.30$ & \citet{Mayor2011}, \citet{SWEET} \\
HD 134606 c & $ 9.26 \pm 0.42 $ & $ 0.27 \pm 0.02 $ & $12.10$ & \citet{Mayor2011}, \citet{SWEET} \\
HD 134606 d & $ 5.20 \pm 0.58 $ & $ 0.27 \pm 0.02 $ & $26.90$ & \citet{Mayor2011}, \citet{SWEET} \\
HD 134606 e & $ 10.70 \pm 0.76 $ & $ 0.27 \pm 0.02 $ & $58.80$ & \citet{Mayor2011}, \citet{SWEET} \\
HD 134606 f & $ 6.90 \pm 1.20 $ & $ 0.27 \pm 0.02 $ & $147.50$ & \citet{Mayor2011}, \citet{SWEET} \\
HD 136352 b & $ 5.28 \pm 0.62 $ & $ -0.34 \pm 0.01 $ & $11.56$ & \citet{Mayor2011}, \citet{SWEET} \\
HD 136352 c & $ 11.38 \pm 0.10 $ & $ -0.34 \pm 0.01 $ & $27.60$ & \citet{Mayor2011}, \citet{SWEET} \\
HD 136352 d & $ 9.59 \pm 1.86 $ & $ -0.34 \pm 0.01 $ & $106.70$ & \citet{Mayor2011}, \citet{SWEET} \\
HD 13808 b & $ 10.33 \pm 0.92 $ & $ -0.21 \pm 0.02 $ & $14.20$ & \citet{Mayor2011}, \citet{SWEET} \\
HD 13808 c & $ 11.55 \pm 1.62 $ & $ -0.21 \pm 0.02 $ & $53.80$ & \citet{Mayor2011}, \citet{SWEET} \\
HD 1461 b & $ 6.44 \pm 0.61 $ & $ 0.19 \pm 0.01 $ & $13.50$ & \citet{Diaz2015}, \citet{SWEET} \\
HD 1461 c & $ 5.92 \pm 0.76 $ & $ 0.19 \pm 0.01 $ & $13.50$ & \citet{Diaz2015}, \citet{SWEET} \\
HD 154088 b & $ 6.15 \pm 0.86 $ & $ 0.28 \pm 0.03 $ & $18.60$ & \citet{Mayor2011}, \citet{SWEET} \\
HD 156668 b & $ 4.15 \pm 0.59 $ & $ -0.04 \pm 0.05 $ & $4.65$ & \citet{Howard2011a}, \citet{SWEET} \\
HD 157172 b & $ 38.10 \pm 2.60 $ & $ 0.11 \pm 0.02 $ & $104.80$ & \citet{Mayor2011}, \citet{SWEET} \\
HD 16417 b & $ 21.28 \pm 1.89 $ & $ 0.13 \pm 0.01 $ & $17.24$ & \citet{OToole2009a}, \citet{SWEET} \\
HD 164595 b & $ 16.14 \pm 2.72 $ & $ -0.04 \pm 0.08 $ & $40.00$ & \citet{Courcol2015}, \citet{deMello2014} \\
HD 179079 b & $ 27.50 \pm 2.50 $ & $ 0.27 \pm 0.02 $ & $14.48$ & \citet{Valenti2009}, \citet{SWEET} \\
HD 181433 b & $ 7.54 \pm 0.68 $ & $ 0.36 \pm 0.18 $ & $9.37$ & \citet{Bouchy2009}, \citet{SWEET} \\
HD 189567 b & $ 8.46 \pm 0.59 $ & $ -0.24 \pm 0.01 $ & $14.30$ & \citet{Mayor2011}, \citet{SWEET} \\
HD 189567 c & $ 7.23 \pm 0.85 $ & $ -0.24 \pm 0.01 $ & $33.62$ & \citet{Mayor2011}, \citet{SWEET} \\
HD 189567 d & $ 7.40 \pm 1.40 $ & $ -0.24 \pm 0.01 $ & $61.72$ & \citet{Mayor2011}, \citet{SWEET} \\
HD 190360 c & $ 18.74 \pm 2.12 $ & $ 0.24 \pm 0.05 $ & $17.11$ & \citet{Wright2009}, \citet{SWEET} \\
HD 192310 b & $ 16.90 \pm 0.90 $ & $ -0.03 \pm 0.04 $ & $74.72$ & \citet{Pepe2011}, \citet{SWEET} \\
HD 20003 b & $ 12.00 \pm 0.97 $ & $ 0.04 \pm 0.02 $ & $11.90$ & \citet{Mayor2011}, \citet{SWEET} \\
HD 20003 c & $ 13.42 \pm 1.28 $ & $ 0.04 \pm 0.02 $ & $33.80$ & \citet{Mayor2011}, \citet{SWEET} \\
HD 204313 c & $ 17.6 \pm 1.7 $ & $ 0.18 \pm 0.02 $ & $34.90$ & \citet{Diaz2015}, \citet{SWEET} \\
HD 20781 b & $ 3.83 \pm 0.73 $ & $ -0.11 \pm 0.02 $ & $5.30$ & \citet{Mayor2011}, \citet{SWEET} \\
HD 20781 c & $ 5.51 \pm 0.47 $ & $ -0.11 \pm 0.02 $ & $13.90$ & \citet{Mayor2011}, \citet{SWEET} \\
HD 20781 d & $ 10.63 \pm 0.63 $ & $ -0.11 \pm 0.02 $ & $29.20$ & \citet{Mayor2011}, \citet{SWEET} \\
HD 20781 e & $ 15.30 \pm 0.81 $ & $ -0.11 \pm 0.02 $ & $85.40$ & \citet{Mayor2011}, \citet{SWEET} \\
HD 20794 b & $ 2.70 \pm 0.31 $ & $ -0.40 \pm 0.01 $ & $18.32$ & \citet{Pepe2011}, \citet{SWEET} \\
HD 20794 c & $ 2.36 \pm 0.43 $ & $ -0.40 \pm 0.01 $ & $40.11$ & \citet{Pepe2011}, \citet{SWEET} \\
HD 20794 d & $ 4.70 \pm 0.57 $ & $ -0.40 \pm 0.01 $ & $90.31$ & \citet{Pepe2011}, \citet{SWEET} \\
HD 215152 b & $ 2.78 \pm 0.47 $ & $ -0.08 \pm 0.02 $ & $7.28$ & \citet{Mayor2011}, \citet{SWEET} \\
HD 215152 c & $ 10.00 \pm 0.48 $ & $ -0.08 \pm 0.02 $ & $10.87$ & \citet{Mayor2011}, \citet{SWEET} \\
HD 215456 b & $ 32.21 \pm 2.92 $ & $ -0.09 \pm 0.01 $ & $193.00$ & \citet{Mayor2011}, \citet{SWEET} \\
HD 215497 b & $ 6.63 \pm 0.79 $ & $ 0.25 \pm 0.05 $ & $3.93$ & \citet{LoCurto2010}, \citet{SWEET} \\
HD 21693 b & $ 10.22 \pm 1.46 $ & $ 0.00 \pm 0.02 $ & $22.70$ & \citet{Mayor2011}, \citet{SWEET} \\
HD 21693 c & $ 20.57 \pm 1.80 $ & $ 0.00 \pm 0.02 $ & $53.90$ & \citet{Mayor2011}, \citet{SWEET} \\
HD 219828 b & $ 19.77 \pm 1.56 $ & $ 0.19 \pm 0.03 $ & $3.83$ & \citet{Melo2007}, \citet{SWEET} \\
HD 31527 b & $ 11.55 \pm 0.80 $ & $ -0.17 \pm 0.01 $ & $16.50$ & \citet{Mayor2011}, \citet{SWEET} \\
HD 31527 c & $ 15.82 \pm 1.10 $ & $ -0.17 \pm 0.01 $ & $51.30$ & \citet{Mayor2011}, \citet{SWEET} \\
HD 31527 d & $ 16.50 \pm 3.00 $ & $ -0.17 \pm 0.01 $ & $274.50$ & \citet{Mayor2011}, \citet{SWEET} \\
HD 38858 b & $ 12.43 \pm 1.70 $ & $ -0.22 \pm 0.01 $ & $198.00$ & \citet{Mayor2011}, \citet{SWEET} \\
HD 39194 b & $ 3.72 \pm 0.33 $ & $ -0.61 \pm 0.02 $ & $5.63$ & \citet{Mayor2011}, \citet{SWEET} \\
HD 39194 c & $ 5.94 \pm 0.47 $ & $ -0.61 \pm 0.02 $ & $14.03$ & \citet{Mayor2011}, \citet{SWEET} \\
HD 39194 d & $ 5.14 \pm 0.66 $ & $ -0.61 \pm 0.02 $ & $33.90$ & \citet{Mayor2011}, \citet{SWEET} \\
HD 40307 b & $ 3.81 \pm 0.3 $ & $ -0.36 \pm 0.02 $ & $4.31$ & \citet{Diaz2015}, \citet{SWEET} \\
HD 40307 c & $ 6.43 \pm 0.44 $ & $ -0.36 \pm 0.02 $ & $9.62$ & \citet{Diaz2015}, \citet{SWEET} \\
HD 40307 d & $ 8.74 \pm 0.58 $ & $ -0.36 \pm 0.02 $ & $20.42$ & \citet{Diaz2015}, \citet{SWEET} \\
HD 40307 e & $ 3.52 \pm 0.13 $ & $ -0.36 \pm 0.02 $ & $34.62$ & \citet{Tuomi2013}, \citet{SWEET} \\
HD 40307 f & $ 3.63 \pm 0.6 $ & $ -0.36 \pm 0.02 $ & $51.56$ & \citet{Diaz2015}, \citet{SWEET} \\
HD 40307 g & $ 7.10 \pm 0.90 $ & $ -0.36 \pm 0.02 $ & $197.80$ & \citet{Tuomi2013}, \citet{SWEET} \\
HD 4308 b & $ 13.00 \pm 1.40 $ & $ -0.34 \pm 0.01 $ & $15.56$ & \citet{OToole2009b}, \citet{SWEET} \\
\hline
\end{tabular}
\end{table}

\newpage
\begin{table}
\centering
\contcaption{Annex : Parameters of the sample}
\begin{tabular}{lcccc}
\hline
Name & Mass [\Mearth] & [Fe/H] & Period [days] & References \\
\hline
HD 45184 b & $ 11.32 \pm 0.83 $ & $ 0.04 \pm 0.01 $ & $5.90$ & \citet{Mayor2011}, \citet{SWEET} \\
HD 45184 c & $ 8.98 \pm 1.13 $ & $ 0.04 \pm 0.01 $ & $13.13$ & \citet{Mayor2011}, \citet{SWEET} \\
HD 47186 b & $ 22.63 \pm 0.88 $ & $ 0.23 \pm 0.02 $ & $4.08$ & \citet{Bouchy2009}, \citet{SWEET} \\
HD 49674 b & $ 32.28 \pm 2.61 $ & $ 0.33 \pm 0.06 $ & $4.95$ & \citet{Butler2006}, \citet{SWEET} \\
HD 51608 b & $ 13.14 \pm 0.98 $ & $ -0.07 \pm 0.01 $ & $14.10$ & \citet{Mayor2011}, \citet{SWEET} \\
HD 51608 c & $ 17.97 \pm 2.61 $ & $ -0.07 \pm 0.01 $ & $95.42$ & \citet{Mayor2011}, \citet{SWEET} \\
HD 69830 b & $ 10.06 \pm 0.55 $ & $ -0.06 \pm 0.02 $ & $8.67$ & \citet{Lovis2006}, \citet{SWEET} \\
HD 69830 c & $ 11.69 \pm 0.81 $ & $ -0.06 \pm 0.02 $ & $31.56$ & \citet{Lovis2006}, \citet{SWEET} \\
HD 69830 d & $ 17.90 \pm 1.66 $ & $ -0.06 \pm 0.02 $ & $197.00$ & \citet{Lovis2006}, \citet{SWEET} \\
HD 7924 b & $ 8.68 \pm 0.52 $ & $ -0.22 \pm 0.04 $ & $5.40$ & \citet{Fulton2015}, \citet{SWEET} \\
HD 7924 c & $ 7.86 \pm 0.72 $ & $ -0.22 \pm 0.04 $ & $15.30$ & \citet{Fulton2015}, \citet{SWEET} \\
HD 7924 d & $ 6.44 \pm 0.79 $ & $ -0.22 \pm 0.04 $ & $24.45$ & \citet{Fulton2015}, \citet{SWEET} \\
HD 85512 b & $ 3.62 \pm 0.44 $ & $ -0.26 \pm 0.14 $ & $58.43$ & \citet{Pepe2011}, \citet{SWEET} \\
HD 90156 b & $ 17.97 \pm 1.49 $ & $ -0.24 \pm 0.01 $ & $49.77$ & \citet{Mordasini2011}, \citet{SWEET} \\
HD 93385 b & $ 4.02 \pm 0.48 $ & $ 0.02 \pm 0.01 $ & $7.34$ & \citet{Mayor2011}, \citet{SWEET} \\
HD 93385 c & $ 7.20 \pm 0.58 $ & $ 0.02 \pm 0.01 $ & $13.18$ & \citet{Mayor2011}, \citet{SWEET} \\
HD 93385 d & $ 7.78 \pm 0.87 $ & $ 0.02 \pm 0.01 $ & $45.84$ & \citet{Mayor2011}, \citet{SWEET} \\
HD 96700 b & $ 9.08 \pm 0.41 $ & $ -0.18 \pm 0.01 $ & $8.13$ & \citet{Mayor2011}, \citet{SWEET} \\
HD 96700 c & $ 3.22 \pm 0.56 $ & $ -0.18 \pm 0.01 $ & $19.90$ & \citet{Mayor2011}, \citet{SWEET} \\
HD 96700 d & $ 12.25 \pm 0.98 $ & $ -0.18 \pm 0.01 $ & $103.22$ & \citet{Mayor2011}, \citet{SWEET} \\
HD 97658 b & $ 7.55 \pm 0.80 $ & $ -0.35 \pm 0.02 $ & $9.49$ & \citet{VanGrootel2014}, \citet{SWEET} \\
HD 99492 b & $ 33.75 \pm 3.72 $ & $ 0.24 \pm 0.12 $ & $17.04$ & \citet{Butler2006}, \citet{SWEET} \\
HIP 116454 b & $ 11.82 \pm 1.33 $ & $ -0.12 \pm 0.03 $ & $9.12$ & \citet{Vanderburg2015}, \citet{SWEET} \\
HIP 57274 b & $ 11.62 \pm 1.31 $ & $ 0.01 \pm 0.06 $ & $8.14$ & \citet{Fischer2012}, \citet{SWEET} \\
Kepler-10 c & $ 17.20 \pm 1.90 $ & $ -0.15 \pm 0.04 $ & $45.30$ & \citet{Dumusque2014}, \citet{SWEET} \\
Kepler-11 d & $ 7.30 \pm 1.00 $ & $ 0.00 \pm 0.10 $ & $22.70$ & \citet{Lissauer2013}, \citet{SWEET} \\
Kepler-18 c & $ 18.40 \pm 2.70 $ & $ 0.20 \pm 0.04 $ & $7.64$ & \citet{HaddenLithwick2014}, \citet{SWEET} \\
Kepler-18 d & $ 15.70 \pm 2.00 $ & $ 0.20 \pm 0.04 $ & $14.86$ & \citet{HaddenLithwick2014}, \citet{SWEET} \\
Kepler-307 b & $ 3.10 \pm 0.60 $ & $ 0.16 \pm 0.15 $ & $10.42$ & \citet{Xie2014}, MAST catalog \\
Kepler-36 b & $ 4.45 \pm 0.33 $ & $ -0.20 \pm 0.06 $ & $13.84$ & \citet{Carter2012}, \citet{SWEET} \\
Kepler-36 c & $ 8.08 \pm 0.60 $ & $ -0.20 \pm 0.06 $ & $16.24$ & \citet{Carter2012}, \citet{SWEET} \\
Kepler-48 c & $ 14.61 \pm 2.30 $ & $ 0.17 \pm 0.07 $ & $9.67$ & \citet{Marcy2014} \\
Kepler-51 c & $ 4.00 \pm 0.40 $ & $ -0.08 \pm 0.15 $ & $85.30$ & \citet{Masuda2014}, MAST catalog \\
Kepler-56 b & $ 22.10 \pm 3.70 $ & $ 0.20 \pm 0.16 $ & $10.50$ & \citet{Huber2013} \\
Kepler-89 e & $ 13.00 \pm 2.50 $ & $ -0.01 \pm 0.04 $ & $54.30$ & \citet{Masuda2013}, \citet{Hirano2012} \\
Kepler-93 b & $ 4.02 \pm 0.68 $ & $ -0.18 \pm 0.10 $ & $4.73$ & \citet{Dressing2015} \\
KOI-620.02 & $ 7.60 \pm 1.10 $ & $ -0.08 \pm 0.15 $ & $130.20$ & \citet{Masuda2014}, MAST catalog \\
mu Ara c & $ 10.50 \pm 0.50 $ & $ 0.32 \pm 0.04 $ & $9.63$ & \citet{Pepe2007}, \citet{SWEET} \\
mu Ara d & $ 10.99 \pm 0.63 $ & $ 0.32 \pm 0.04 $ & $9.64$ & \citet{Pepe2007}, \citet{SWEET} \\
Kapteyn's c & $ 7.00 \pm 1.10 $ & $ -0.89 \pm 0.15 $ & $121.54$ & \citet{Anglada2014} \\
HD 219134 b & $ 4.46 \pm 0.47 $ & $ 0.11 \pm 0.04 $ & $3.09$ & \citet{Motalebi2015} \\
HD 219134 d & $ 8.67 \pm 1.14 $ & $ 0.11 \pm 0.04 $ & $46.78$ & \citet{Motalebi2015} \\
HD 219134 f & $ 8.90 \pm 1.00 $ & $ 0.11 \pm 0.04 $ & $22.80$ & \citet{Vogt2015}, \citet{Motalebi2015} \\
HD 219134 g & $ 11.00 \pm 1.00 $ & $ 0.11 \pm 0.04 $ & $94.20$ & \citet{Vogt2015}, \citet{Motalebi2015} \\
HD 175607 b & $ 8.98 \pm 1.1 $ & $ -0.62 \pm 0.01 $ & $29.01$ & \citet{Mortier2016} \\
\hline
\end{tabular}
\end{table}


\bsp	
\label{lastpage}
\end{document}